\documentstyle[psfig]{mn}
\newcommand{\etal}{{\rm {et al.}$\:$}}
\newcommand{\gesim}{\,\raisebox{-0.4ex}{$\stackrel{>}{\scriptstyle\sim}$}\,}
\newcommand{\lesim}{\,\raisebox{-0.4ex}{$\stackrel{<}{\scriptstyle\sim}$}\,}
\newcommand{\lskip}{\vskip \baselineskip}
\newcommand{\nskip}{\lskip \noindent}
\newcommand{\En}{\mbox{$E_{\rm 20}$}}
\newcommand{\rg}{\mbox{$r_{\rm g}(E)$}}
\newcommand{\rsq}{\mbox{$r_{\rm g}^{2}(E)$}}

\newcommand{\Bng}{\mbox{$B_{\rm -9}$}}
\newcommand{\lc}{\mbox{$\ell_{\rm coh}$}}
\newcommand{\lcc}{\mbox{$\ell_{\rm c}$}}
\newcommand{\mpr}{\mbox{$m_{\rm p}$}}

\newcommand{\lE}{\mbox{$\ell(E)$}}

\newcommand{\cR}{\mbox{${\cal R}$}}

\newcommand{\mpi}{\mbox{$m_{\pi}$}}
\newcommand{\Et}{\mbox{$\overline{E}_{\rm t}$}}
\newcommand{\mpl}{\mbox{$m_{+}$}}
\newcommand{\mmi}{\mbox{$m_{-}$}}
\newcommand{\mpm}{\mbox{$m_{\pm}$}}

\newcommand{\rat}{\mbox{$\displaystyle 
     \left( \frac{\En}{ \mbox{$Z$} \Bng} \right)$}}
\newcommand{\Ep}{\mbox{$E_{\rm p}$}}
\newcommand{\gp}{\mbox{$\gamma_{\rm p}$}}

\newcommand{\bp}{\mbox{$\beta_{\rm p}$}}
\newcommand{\eprf}{\mbox{$\epsilon_{0}$}}
\newcommand{\lav}{\left< \left<}
\newcommand{\rav}{\right> \right>}
\newcommand{\eave}[1]{\left< #1 \right>}
\newcommand{\grad}{\mbox {\boldmath $\nabla$}}
\newcommand{\Etr}{\mbox{$E_{\rm th}$}}
\newcommand{\bm}[1]{\mbox{\boldmath $#1$}}
\newcommand{\bdot}{\mbox{${\bf  \: \cdot \:}$}}
\newcommand{\btimes}{\mbox{${\bf \: \times \:}$}}

\newcommand{\hatn}{\hat{\bm{n}}}
\newcommand{\hatb}{\hat{\bm{b}}}
\newcommand{\evec}[1]{\mbox{$\bm{e}_{\rm #1}$}}

\newcommand{\lpg}{\ell_{\rm p \gamma}}

\newcommand{\half}{\mbox{$\frac{1}{2}$}}
\newcommand{\third}{\mbox{$\frac{1}{3}$}}
 
\begin{document}
 \title{INTERGALACTIC PROPAGATION OF UHE COSMIC RAYS}
 \author[A. Achterberg et al.]
 {Abraham Achterberg$^{1,2}$, 
  Yves Gallant$^{1,3}$, 
  Colin A.\ Norman$^{4,5}$ and 
 Donald B.\ Melrose$^6$\\
 $^1$Astronomical Institute, Utrecht University,
            P.O.\ box 80.000, 3508 TA Utrecht, The Netherlands \\
 $^2$Center for High Energy Astrophysics, Kruislaan 403, 
           1098 SJ Amsterdam, The Netherlands\\
 $^3$Dublin Institute for Advanced Studies, 5 Merrion Square, Dublin 2,
      		Republic of Ireland\\	            
 $^4$Department of Physics and Astronomy, Johns Hopkins University,
 	  Homewood campus, Baltimore MD 21218\\
 $^5$Space Telescope Science Institute, 3700 San Martin Drive,
 	Baltimore, MD 21218\\
 $^6$Research Center for Theoretical Astrophysics, School of Physics,
 University of Sydney, NSW 2006, Australia}		            

\maketitle

 %%%%%%%%%%%%%%%%%%%%%%%%%%%%%%%%%%%%%%%%%%%%%%%%%%%%%%%%%%%%%%%%%%%%%%%%%%%
 \begin{abstract}
 We discuss the intergalactic propagation of ultra-high-energy cosmic rays 
 (UHECRs) with energies $E \gesim 10^{18.5}$ eV. We consider the propagation 
 of UHECRs under the influence of the energy-dependent deflection by a weak 
 random magnetic field in the intergalactic medium and energy losses by photo-pion 
 and pair production. We calculate arrival spectra taking full account of the 
 kinematics of photo-pion production and the Poisson statistics of the photo-pion 
 interaction rate.
 
 We give estimates for the deflection of UHECRs from the line of sight
 to the source, time delays with respect to photons from the same source,
 arrival spectra and source statistics. These estimates are confirmed by
 numerical simulations of the propagation in energy evolution of UHECRs.
 These simulations demonstrate that the often-used continuous approximation
 in the treatment of energy losses due to photo-pion production on the
 cosmic microwave background (CMWB) cannot be justified for UHECRs.
 
 We discuss the implications of these results for the observed flux of
 particles above the Greisen-Zatsepin-Kuz'min cut-off in the two of the scenarios
 that have been proposed for the production of these particles: continuous
 production in the large shock waves associated with powerful radio galaxies,
 or possibly large-scale structure formation, and the impulsive production at 
 relativistic blast waves associated with cosmological gamma-ray bursts.
 
 \end{abstract}


 \section{Introduction}
 
Recent observations of ultra-high energy cosmic rays (UHECRs) above 
$10^{18.5}$ eV (Bird \etal, 1994, Yoshida \etal, 1994, Takeda \etal, 1998) 
suggest that these particles are of extragalactic origin. 
The indications are threefold:
\begin{enumerate}
\item 
The energy spectrum hardens, from $J(E) \propto E^{-3.2 \pm 0.1}$ below 
$10^{18.5}$ eV to $J(E) \propto E^{-2.8 \pm 0.3}$ above $10^{18.5}$ eV; 
\item 
Arrival directions do not seem to be clustered significantly along the 
Galactic plane, although there is an indication that they may be clustered along 
the supergalactic plane for $E > 10^{19.5}$ eV
(Hayashida \etal, 1996; Stanev \etal, 1995). 
\item 
There is a suggestion that the composition changes from one 
dominated by heavy nuclei (iron group) below $10^{18.5}$ eV, as expected 
for a population of Galactic origin, to one dominated by light nuclei or 
even protons above $10^{18.5}$ eV; 
\end{enumerate}

At these energies the deflection of UHECRs in the $\sim \: \mu$G 
Galactic magnetic fields is too small to randomise flight directions within 
the Galaxy.  In terms of the typical gyration radius of a particle with energy 
$E$ and charge $q = Ze$ and pitch angle $i$, 
$\rho_{\rm G} = \rg \sin{i}$ with  $\rg \equiv E/ZeB$, the deflection 
angle $\Delta \theta$ inside the Galactic disk of size $d \sim 10$ kpc is 
of order
\begin{equation}
	\Delta \theta \simeq \frac{d}{\rg} =
	6^{o} \: (Z B_{\mu {\rm G}}/E_{20})  \; ,
\end{equation}
with $E_{20} \equiv E/(10^{20}$ eV). In practice, it will be much less
due to the inclination of the particle orbit with respect to the disk plane 
and the fact that the Galactic magnetic field is strongly turbulent
on kpc scales (for a review: Zweibel \& Heiles, 1997), and will not act
coherently over the whole disk. Therefore, an origin of UHECRs within the Galactic disk
should lead to a significant clustering of arrival directions
around the Galactic plane, unless the sources are in an extensive halo around the
Galaxy.

From a theoretical point of view, the fundamental limit on the maximum
energy attainable in any electrodynamic acceleration mechanism relying on 
bulk plasma motions (i.e. waves or shocks) is
(Norman, Melrose \& Achterberg, 1995, Gallant \& Achterberg, 1999)
\begin{equation}
	E_{\rm max} \sim Ze \Gamma_{\rm s} 
	\beta_{\rm s} B R_{\rm s} \; .
\end{equation}
Here $\beta_{\rm s} = V/c$ is the wave or bulk velocity in units of
the speed of light, $\Gamma_{\rm s} = (1 - \beta_{\rm s}^{2})^{-1/2}$
is the associated Lorentz factor, $B$ is the magnetic field strength 
and $R_{\rm s}$ is the linear size of the source. This limit strongly 
constrains any model for
the production of UHECRs (Norman, Melrose \& Achterberg, 1995).
Galactic sources, with the possible exception of radio pulsars, seem
to be excluded. The value of $E_{\rm max}$ for these source falls well
below the UHECR energy range.

For the above reasons, most theoretical discussions of the origin of these
particles have centered on extragalactic sources. Leaving aside for the 
moment the more exotic models involving some fundamental particle or its
decay products (e.g. Farrar \& Biermann, 1998) or quantum-mechanical topological 
defects (such as superconducting strings, e.g. Sigl \etal, 1994), 
there are two major classes of models describing the production of UHECRs:
\begin{itemize}
\item	{\bf Continuous production} of UHECRs in the kpc-scale shocks 
          associated with the jets of powerful Fanaroff-Riley class II 
          radio galaxies (Rachen \& Biermann, 1993; 
          Norman, Melrose \& Achterberg, 1995), 
          or in shocks associated with ongoing large-scale structure 
          formation in clusters of galaxies (Norman, Melrose \& Achterberg, 1995; 
          Kang, Ryu \& Jones, 1996; Kang, Rachen \& Biermann, 1997);
           
\item	{\bf Impulsive production} in the same sources responsible for
	gamma-ray bursts (Waxman, 1995a; Vietri, 1995; Milgrom \& Usov, 1995).
	This possibility has been proposed  on the basis of a striking coincidence: 
	the typical UHECR energy flux above $10^{19}$ eV, and the typical
	gamma-ray flux from gamma-ray bursters in the 
	cosmological scenario (Pacz\'ynski, 1986) is roughly the same.
	The recent identification of optical counterparts for several
	gamma-ray bursts (van Paradijs \etal, 1997; Djorgovski \etal, 1997;
	Metzger \etal, 1997) seems to indicate that they indeed occur at
	cosmological distances, and not in an extended halo around our own 
	galaxy as had also been proposed. 
\end{itemize}
A recent review of the various issues involved in the production and propagation
of UHECRs can be found in Biermann (1995).

A complication in all these models is the limited volume in the local universe, 
corresponding to a typical radius $D_{\rm max} \sim 50$ Mpc, from which 
UHRCRs above $10^{19.5}$ eV must originate if they are indeed protons or light nuclei, 
and not some exotic particle. Interaction between photons of the Cosmic Microwave 
Background (CMWB) and UHECRs leads to photo-pion production, 
e.g. $p + \gamma \longrightarrow p + \pi$'s. This process strongly degrades the energy of 
particles above the Greisen-Zatsepin-Kuz'min cut-off energy $E_{\rm c} \simeq
10^{19.5}$ eV (Greisen, 1966; Zatsepin \& Kuz'min, 1966) with a typical loss length
of about 20-30 Mpc above $10^{20}$ eV. 

The observation in 1991 with the Fly's Eye detector of a particle with an 
energy of $E \simeq 3 \times 10^{20}$ eV (Bird \etal, 1995), and more recently of six
events above $10^{20}$ eV with the Akeno Giant Air Shower Array (Takeda \etal, 1998),
all well above the GZK cut-off, suggests either  relatively close-by but unidentified
continuous sources ($D < 50$ Mpc, e.g. Elbert \& Sommers, 1995), or the scenario
where particles are produced impulsively and arrive with a strongly energy-dependent
delay with respect to photons due to scattering on weak intergalactic magnetic 
fields (Waxman, 1995a,b; Waxman \& Coppi, 1996; Miralda-Escud\'e \& Waxman, 1996; 
Waxman \& Miralda-Escud\'e, 1996). In the latter case, an identification of the
source in other wavelength bands is not to be expected if the delay
is large enough. As we will show, for reasonable values of the magnetic field strength
in the intergalactic medium, the delay of UHECRs with respect to photons could be
as large as $10^{3}-10^{5}$ years.     

In this paper we will not address the production mechanism of UHECRs. 
Rather we will concentrate on the propagation of these particles
in the intergalactic medium (IGM) under the influence of weak
IGM magnetic fields and their interaction with the cosmic microwave
background. Scattering on IGM magnetic fields leads to  diffusion in
the flight direction of UHECR particles, which induces
a time delay with respect to photons emitted simultaneously at the 
source (e.g. Waxman, 1995; Miralda-Escud\'e \& Waxman, 1996). 
The latter effect is especially important in GRB models of UHECR production,
where a time delay between the primary gamma ray photons and `secondary' UHECRs
produced concurrently may yield a strong observational constraint on
the strength of the IGM magnetic field.
Both the time delay and rms deflection are strong functions of particle
energy, $t_{\rm del} \propto E^{-2}$ and $\alpha_{\rm rms} \propto E^{-1}$,
so that losses incurred by UHECR particles due to their interaction
with the CMWB (pion production, and pair production below $10^{19}$ eV) 
strongly influence particle propagation.

In Section 2 we systematically derive the relevant equations for the 
propagation of UHECRs in a weak, random intergalactic magnetic field. 
In Section 3, the precise form of the energy losses is considered.
We re-examine the treatment of energy losses of UHECRs and show
that a continuous approximation in terms of the mean energy loss is not a 
good approximation for particles above $10^{19}$ eV.  
In particular we will show that the
correct treatment of the Poisson statistics of photon-UHECR encounters
leads to a significant high-energy tail in the spectrum for sources
closer than $\sim 50$ Mpc.
In Section 4 we present numerical simulations of cosmic-ray propagation.
The observational consequences in terms of the different UHECR production
scenarios are discussed in  Section 5. Conclusions are presented in Section 6.

\section{UHECR propagation in intergalactic fields}

In the tenuous intergalactic medium (IGM), magnetic fields are responsible for the
deflection of UHECRs. Direct measurements of the IGM field strength 
{\em outside} clusters are not available. Observational limits (Kronberg, 1996) 
can be placed on the ordered (large-scale) component 
$\left<B \right>_{\rm IGM}$ 
and the amplitude $B_{\rm r}$ of the random component of 
the intergalactic magnetic field.
The random component is assumed to have a coherence length
$\lc \simeq 1$ Mpc. These limits are:
\begin{eqnarray}
\label{Blimits}
	\left< B \right>_{\rm IGM} & \lesim & 10^{-11} \; {\rm G} \; , 
	\nonumber \\
	& & \\
	B_{\rm r} & \lesim & 10^{-9} \; {\rm G} .
	\nonumber
\end{eqnarray}
The field strengths within clusters of galaxies can be much larger,
$B \simeq 10^{-7}-10^{-6}$ G.

The typical gyration radius of an ultra-relativistic cosmic ray 
particle with charge $q = Ze$ and energy $E$ in a magnetic field
$B$ is
\begin{equation}
	\rg \simeq  \frac{E}{ZeB} \simeq 0.1 \rat \; {\rm Gpc} \; .
\end{equation}	 
Here we have defined
\begin{equation}
	\En \equiv \frac{E}{10^{20} \; {\rm eV}} \; \; \; , \; \; \; 
	\Bng \equiv \frac{B}{10^{-9} \; {\rm G}} \; .
\end{equation}
These estimates imply that the turning angle of UHECR particles in the
average field $\left< B \right>_{\rm IGM}$ over a distance $D$ will be
\begin{equation}
	\theta \simeq \frac{D}{\rg} \simeq
	0.53^{o}  \: \frac{Z \: D_{2}}{E_{20}}  \: 
	\left( \frac{\left< B \right>_{\rm IGM}}{10^{-11} \; {\rm G}}\right)
	\; .
\end{equation}
In this expression we defined $D_{2} \equiv D/(100 \; {\rm Mpc})$.
This estimate assumes a uniform field so it is actually an upper limit.

The time delay with respect to ballistic propagation
follows from expanding the relationship between the
path length $\ell$ along the circular orbit of a charged particle
with gyration radius $\rg$ and the linear distance $D$ travelled.
It equals, for $\ell \ll \rg$,
\begin{equation}
	t_{\rm delay} \simeq \frac{D^{3}}{24 c \: \rsq} \simeq
	1200 \; \frac{Z^{2}D_{2}^{3}}{E_{20}^{2}} \: 
	\left( \frac{\left< B \right>_{\rm IGM}}{10^{-11} \; {\rm G}}\right)^{2}
	\; {\rm yr} \; .
\end{equation}
In what follows, we will assume that the influence of the mean field on 
UHECR propagation can be neglected with respect to the action of the 
random field component. We do not believe that a the intergalactic 
magnetic field will be uniform  on scales exceeding $\sim 10$ Mpc, the
typical scale associated the large-scale structure of our Universe.
In that case, the contribution of the large-scale  field to the
delay would be measured in terms of a few years around $10^{20}$ eV.

\subsection{Small-angle scattering by random fields}

The equations of motion for an ultra-relativistic particle with charge
$Ze$ and energy $E$ in a quasi-static magnetic field can be written as
\begin{equation}
\label{eom}
	\frac{{\rm d} \bm{x}}{{\rm d} s} = \hatn \; \; \; , \; \; \; 
	\frac{{\rm d} \hatn}{{\rm d} s}  = 
	\left( \frac{Ze|\bm{B}|}{E} \right) \: 
	\hatn \btimes \hatb \; .
\end{equation}
Here $s \equiv ct$ is the path length along the orbit, 
$\hatn \equiv (n_{1} \: , \: n_{2} \: , \: n_{3})$ is the unit vector along
the direction of flight and $\hatb \equiv (b_{1} \: , \: b_{2} \: , \: b_{3})$ 
is the unit vector along the magnetic field.
For simplicity we will assume that the magnetic field has a uniform amplitude
$B_{\rm r}$ and a random (isotropic) direction distribution so that
\begin{equation}
\label{cells}
	|\bm{B}|  = B_{\rm r} \; \; , \; \; 
	\lav b_{i} \rav = 0 \; \; , \; \; 
	\lav b_{i} b_{j} \rav = \mbox{$\frac{1}{3}$} \: \delta_{ij} \; .
\end{equation}
Here the double brackets $\lav \cdots \rav$ denote an ensemble average.
This assumption is equivalent to a collection of randomly oriented cells
with a uniform value of the magnetic field strength.
The quasi-static treatment of the magnetic field corresponds to the
requirement
\begin{equation}
	\left|( \hatn \bdot \grad) \hatb \right|  
	\gg \left| \frac{1}{c} \: \frac{\partial \hatb}{\partial t} \right| \; ,
\end{equation}
which is always satisfied as the bulk (turbulent) velocity in the IGM is
much less than the speed of light.

The equation of motion for $\hatn$ can be written in component form as:
\begin{equation}
	\frac{{\rm d}n_{i}}{{\rm d} s} = 
	\frac{\epsilon_{ijk} \: n_{j}b_{k}}{\rg} \; ,
\end{equation}
with $\rg \equiv E/ZeB_{\rm r}$ and $\epsilon_{ijk}$ the totally anti-symmetric 
Levi-Cevita tensor. The summation convention is used here and below.
Magnetic scattering leads to diffusion of the flight direction $\hatn$.
The corresponding diffusion coefficient,
\begin{equation}
	{\cal D}_{ij} \equiv 
	\frac{\eave{\Delta n_{i} \: \Delta n_{j}}}{2 \Delta s} \; ,
\end{equation}
follows from the Kubo-Taylor formula 
(Taylor, 1921; see also: Sturrock, 1994) as
\begin{equation}
\label{KuTay}
	{\cal D}_{ij} = 	
	\frac{1}{r_{\rm g}^{2}(E)} \: 
	\int_{0}^{\infty} {\rm d} s \: \epsilon_{ikl} n_{k}(0) 
	\epsilon_{jqr} n_{q}(s) \: 
	\eave{b_{l}(0) b_{r}(\bm{x}(s))} \: .	
\end{equation}
This diffusion tensor is defined per unit path length, and its components
have the dimension [length]$^{-1}$.
If the deflection of the particle incurred in one correlation length is small 
in  the sense that $|\Delta \hatn | \simeq \lc/\rg \ll 1$,
the flight direction $\hatn$ may be approximated as constant (ballistic motion)
and $n_{k}(0) n_{q}(s) \approx n_{k}(0) n_{q}(0)$.  This essentially 
corresponds to the well-known {\em quasi-linear approximation} of wave-particle
interactions (e.g. Davidson, 1972). Using Eqn. (\ref{cells}) one has
\begin{equation}
	\int_{0}^{\infty} {\rm d} s \: \eave{b_{l}(0) b_{r}(\bm{x}(s))} =
	\frac{\lc}{3} \: \delta_{lr} \; .
\end{equation} 
This integral of the two-point correlation $\eave{b_{l}(0) b_{r}(\bm{x}(s))}$
formally defines the coherence length $\lc$ of the random field.
Using the property $\epsilon_{ijk}\epsilon_{ilm} = \delta_{jl} \delta_{km} -
\delta_{jm} \delta_{kl}$ of the Levi-Cevita tensor, 
the resulting diffusion tensor can be written in terms of a scalar diffusion
coefficient ${\cal D}_{0}$ and a projection tensor 
$\delta_{ij} - n_{i}n_{j}$
onto the plane perpendicular to $\hatn$:
\begin{equation}
\label{ndiff}
	{\cal D}_{ij} = {\cal D}_{0} \: \left( \delta_{ij} - n_{i}n_{j} \right) 
	\; \; \; , \; \; \; 
	{\cal D}_{0} = \frac{\lc}{3 \rsq} \; .
\end{equation}
It is easily checked that this diffusion tensor preserves the unit
norm $\hatn \bdot \hatn = 1$ since $n_{i} \: {\cal D}_{ij} \: n_{j} = 0$.

The expression for the scalar diffusion coefficient 
${\cal D}_{0}$ can be explained in terms of a simple model.
A charged particle crossing a magnetic cell of size $2 \lc$ 
and field strength $|B| \sim B_{\rm r}$ turns through an angle
\begin{equation}
	\delta \alpha \simeq \frac{2 ZeB_{\perp} \lc}{E} \; ,
\end{equation} 
where $B_{\perp}$ is the component of the field perpendicular to the particle 
velocity. Deflections in different cells add up diffusively. 
After encountering $N \sim s/2\lc$ cells in a distance $s$ 
particles have turned an angle $\sim \alpha_{\rm rms}$ corresponding
to $\alpha_{\rm rms}^{2}  =
(s/2\lc) \times \lav (2ZeB_{\perp} \lc/E)^{2} \rav$.  The associated
angle diffusion coefficient is related to scalar diffusion coefficient
${\cal D}_{0}$ by $\alpha_{\rm rms}^{2} = 4 {\cal D}_{0}s$  (see Eqn. B.17),
so one finds:
\begin{equation}
	{\cal D}_{0} \equiv
	\frac{\alpha_{\rm rms}^{2}(s)}{4s} \simeq
	\mbox{$\frac{1}{3}$} \: \frac{\lc}{\rsq} \; ,
\end{equation}
Here we used $\lav B_{\perp}^{2} \rav = \mbox{$\frac{2}{3}$} B_{\rm r}^{2}$.

The result (\ref{ndiff}) carries over to the more general case of 
scattering by a broad-band spectrum of random magnetic fields.
As shown in Appendix A, an isotropic distribution of magnetic fields
with a power spectrum ${\cal B}(k)$ as a function of wave number,
normalised in such a way that
\begin{equation}
	B_{\rm rms}^{2} \equiv \eave{| \bm{B} |^{2}} = 
	\int_{0}^{\infty} {\rm d} k \: {\cal B}(k) \; ,
\end{equation} 
leads to a scalar diffusion coefficient of the form
${\cal D}_{0} = \lc/3 \rsq$ with $\rg \equiv E/ZeB_{\rm rms}$ and
an effective correlation length equal to
\begin{equation}
\label{effcorr}
	\lc = \frac{3 \pi}{\displaystyle 8 B_{\rm rms}^{2}} \:   
	\int_{k_{\rm min}}^{\infty} {\rm d} k \: \frac{{\cal B}(k)}{k}	
	\; .
\end{equation}
Here $k_{\rm min} \sim 2 \pi/\rg$ is the smallest wavenumber for which the
approximation of using the unperturbed particle trajectory still
(marginally) applies.

\subsection{Deviation angles and time delays}

In this Section, we consider the deviation angle with respect to the
line of sight to the source, and the time delay with respect to photons
which is accumulated by UHECRs due to small-angle scattering on 
magnetic cells in the intergalactic medium. Detailed calculations 
which consider the influence of scattering on the flight direction and
position of the particles, can be found in Appendix B.

Scattering leads to a deviation angle between the original
direction of flight, taken to be along the $z$-axis, and
the direction of propagation $\hatn$.
Defining $\alpha \equiv {\rm cos}^{-1}(\hatn \bdot \evec{z})$ one
has after a path length $s$:
\begin{equation}
\label{angle1}
	\left< \alpha^{2}\right>(s) \simeq 4 {\cal D}_{0} s = 
	\mbox{$\frac{4}{3}$} \: \frac{s \lc}{r_{\rm g}^{2}(E)} \; . 
\end{equation}
This differs from the naive estimate, $\alpha^{2}(s) \sim 2 {\cal D}_{0} s$,
by a factor of two due to the effect of dynamical friction.

However, the angle between the line of sight from the source, 
$\bm{\hat{r}} \equiv \bm{r}/|\bm{r}|$, and the
particle flight direction as measured by an observer, 
defined by $\alpha' \equiv {\rm cos}^{-1}(\bm{\hat{r}} \bdot \hatn)$,
will be {\em less} due to the statistical 
correlation between the particle position,
$\bm{r} = ( x \: , \: y \: , \: z)$, and the flight direction $\hatn$.
One finds for this angle $\eave{\alpha'} = 0$ and
$(\alpha')^{2}(s) = \frac{4}{3} \: {\cal D}_{0} s$ 
after a path length $s$ (Eqn. \ref{barave}). To zeroth order in ${\cal D}_{0}s$  the
linear distance to the source equals $D \simeq s$ (see below).
An observer at a distance $D$ will therefore measure a spread of
UHECR arrival directions with respect to the local line of sight from
the source equal to
\begin{equation}
\label{rmsangle}
	(\alpha')^{2}(s \simeq D) = 
	\mbox{$\frac{4}{9}$} \: \frac{D \lc}{r_{\rm g}^{2}(E)} \; ,
\end{equation}
which corresponds to and rms value 
$\alpha_{\rm rms} = \sqrt{(\alpha')^{2}}$
equal to
\begin{equation}
\label{alpha rms}
	\alpha_{\rm rms} 
	\sim
	3.5^{o} \:  \left( \frac{Z \Bng}{\En} \right)
	\left( D_{2} \ell_{0} \right)^{1/2} \; .
\end{equation}
Here we defined $\ell_{0} = \lc/(1 \; {\rm Mpc})$ and 
$B_{-9} = B_{\rm r}/(10^{-9} \; {\rm G})$. This estimate {\em neglects}
the changes in particle energy as it propagates. Energy losses are important, 
as we will see in Section 4.

The typical angle between the line of sight to the source from the observer
and the original flight direction, 
$\theta \equiv {\rm cos}^{-1} (\bm{\hat{r}} \bdot \evec{z})$, satisfies
$\left< \theta^{2}(s) \right> = \frac{4}{3} \: {\cal D}_{0} s$, so
its rms value satisfies $\theta_{\rm rms} \simeq \alpha_{\rm rms}$.

Due to scattering, the path of an individual UHECR is corrugated, leading
to a path length increase with respect to ballistic (straight line) propagation.
The average scattering-induced arrival time delay with respect
to ballistic propagation at the speed of light at distance $D$, 
$\left< t_{\rm del}  \right> = (s - D)/c$, is
(Eqn. \ref{propdelay}):
\begin{equation}
\label{delay time}
	\left< t_{\rm del} \right> \simeq
	\left( \frac{\lc}{9c} \right) \: \left( \frac{D}{\rg} \right)^{2} 
\end{equation}
which is
\begin{equation}
\label{del value}	
	\left< t_{\rm del}  \right> \simeq
	3.1 \times 10^{5} \left( \frac{Z \Bng}{\En} \right)^{2} \:
	 D_{2}^{2} \: \ell_{0} \; {\rm yr} \; .  
\end{equation} 
We again assume a constant particle energy.
All these relations are derived with the implicit assumption that
all deviation angles remain small, i.e. ${\cal D}_{0} s \ll 1$.

The deflection angle $\alpha_{\rm rms}$ and the 
arrival delay $\eave{t_{\rm del}}$ should be interpreted with some care. 
Particles originating from the same 
source will only see statistically independent realisations of the random 
magnetic field if they diffuse more than one coherence length apart 
in the  direction perpendicular to the unperturbed orbit. 
The typical perpendicular distance from the original direction of flight at a
path length $s$ from the source equals
\begin{equation}
\label{perpdeviate}
	d_{\perp} \equiv \sqrt{\eave{x^{2}} + \eave{y^{2}}} = 
	\sqrt{\mbox{$\frac{4}{3}$} \: {\cal D}_{0} s^{3}} \; ,
\end{equation}
where we use Eqn. (\ref{smallslim}).	
The requirement $d_{\perp} > \lc$ determines the (energy dependent)
{\em decorrelation distance}:
\begin{equation}
\label{dcorr}
	D_{\rm dc} \simeq \left( \frac{3 \lc^{2}}{4 {\cal D}_{0}}
	\right)^{1/3} = (\mbox{$\frac{3}{2}$})^{2/3} \:  
	\lc^{1/3} r_{\rm g}^{2/3}(E) \; , 
\end{equation}
which is
\begin{equation}
	D_{\rm dc} \simeq 30 \: \ell_{0}^{1/3} \rat^{2/3} \; {\rm Mpc} \; .
\end{equation}
Total decorrelation occurs for $D  \gg D_{\rm dc}$ only.
Particles with energies such that the source distance is less than
$D_{\rm dc}$ {\em will} incur a deviation angle of order 
$\alpha_{\rm rms}$, and an arrival delay
with a typical magnitude $\eave{t_{\rm del}}$, but the dispersion in these
quantities associated with a collection of particles
from the same source is expected to be smaller than for particles
originating at distances much larger than $D_{\rm dc}$, which have traversed 
different magnetic cells. This means that the simulation results of
Section 4 should be interpreted differently for $D \lesim D_{\rm dc}$ and
$D \gg D_{\rm dc}$. In the first case, the distributions are probability
distributions for the time delay, deviation angle etc. averaged over an ensemble
of sources. In the second case,
particles from a single source will be actually distributed according to the
distributions shown.
Given the low fluxes above $10^{18.5}$ eV, it is presently not possible
to measure the resulting dispersion from a single source, and only individual 
arrival directions are measured.

\subsection{Spatial diffusion}

A description of UHECR propagation in terms of {\em spatial}
diffusion only applies for low energies and/or for
large distances. 

At low energies, where $\rg \lesim \lc$, the deflection angle
in traversing a {\em single} cell is already large,
\begin{equation}
	\delta \alpha \simeq \frac{\lc}{\rg} \equiv
	\frac{E_{\ast}}{E} \; ,
\end{equation}
with $E{\ast}$ the characteristic energy for a given field strength  and  
field reversal scale: 
\begin{equation}
	E_{\ast} \equiv 
	ZeB_{\rm r} \: \lc \simeq 9.2 \times 10^{17} Z B_{-9} \ell_{0}
	\; {\rm eV} \; .
\end{equation}
Such particles will diffuse through the network of 
magnetic cells, and presumably within cells if the magnetic turbulence
has a broad band spectrum. To describe their propagation, the details
of the spectrum of magnetic fluctuations becomes important. As a
rough estimate, we assume {\em Bohm diffusion} with a 
mean-free-path $\sim \rg = \lc \: (E/E_{\ast})$ and spatial diffusion 
coefficient ${\cal K}_{\rm B} \equiv c\rg/3$.
In that case the finite age of the Universe, $t \sim H_{0}^{-1}$, limits
the maximum distance of sources contributing to the present-day flux
at $E \lesim E_{\ast}$ to
\begin{eqnarray*}
	D_{\rm diff}(E < E_{\ast}) 
	& \sim &\left(\frac{c \lc}{H_{0}} \right)^{1/2} \:
	\left( \frac{E}{E_{\ast}} \right)^{1/2} \\
	& & \\
	& \simeq &
	55 \; \ell_{0}^{1/2} h^{-1/2} 
	\left( \frac{E}{E_{\ast}} \right)^{1/2}
	\; {\rm Mpc} \; ,
\end{eqnarray*}
where $h = H_{0}/$(100 km s$^{-1}$ Mpc$^{-1}$).

This essentially means that the Universe is not transparent for cosmic
rays with $E < E_{\ast}$ {\em provided} it is filled uniformly with
magnetic cells. This estimate ignores the effects of cosmological evolution. 
These effects are however included  the simulations presented in Section 4.

Assuming that UHECRs are indeed of extragalactic  origin, this
implies an exponential cut-off in the UHECR flux at Earth below 
$E \simeq E_{\ast}$.
For particles with $E > E_{\ast}$ 
the deviation angle accumulated by traversing many cells
becomes large (see Eqn. B22/23) at a distance of order
\begin{equation}
\label{meanfree}
	\lambda(E) = \frac{1}{2 {\cal D}_{0}} = 
	\frac{3 r_{\rm g}^{2}(E)}{2 \lc} \; ,
\end{equation}
which is
\begin{equation}
	\lambda(E) = 15 \; \rat^{2} \ell_{0}^{-1} \; {\rm Gpc}  
\end{equation}
As shown in Appendix B,  the propagation of UHECRs on these scales
can be treated as spatial diffusion with a diffusion coefficient 
equal to (Eqn. \ref{spatcoeff})
\begin{equation}
	{\cal K} \equiv \frac{D^{2}}{2t} = c \lambda(E) \; .	 
\end{equation}
The finite age of the universe, $t \sim H_{0}^{-1}$, gives a  
maximum distance of UHECR
sources contributing to the present-day flux:
\begin{equation}
	D_{\rm diff}(E) \simeq \sqrt{\frac{2 {\cal K}}{H_{0}}} = 
	\left( \frac{3c \lc}{H_{0}} \right)^{1/2} 
	\: 
	\left( \frac{E}{E_{\ast}} \right) \; .
\end{equation}
The rapid energy losses due to photo-pion production above 
$10^{19.5}/(1 + z)$ eV, followed by a much slower loss at lower energies, 
means that effectively one can put
$E \le 10^{19.5}$ eV in these estimates. So for $E > E_{\ast}$,
sources contributing to the UHECR flux must be closer than
\begin{equation}
	D_{\rm diff}(10^{19.5} \; {\rm eV}) \simeq
	600 \; \ell_{0}^{-1/2} h^{-1/2} \: (Z B_{-9})^{-1} \; 
	{\rm Mpc} \; .
\end{equation}

\subsection{Effects of large-scale structure}

So far, we have treated the effects of the intergalactic magnetic
field in a simple model where the field consists of randomly oriented cells
which are distributed more-or-less homogeneously.
Although true observational evidence is lacking, one might argue
that magnetic fields follow the distribution of luminous matter and
are concentrated mainly in the filamentary large-scale structure
of the universe (e.g. de Lapparent \etal, 1986).
A concentration of strong magnetic fields in the filaments
observed in the large-scale structure could be expected, in particular 
if fields are amplified considerably in the turbulence associated with 
the formation of galaxies, as proposed by Kulsrud \etal (1997). 
Such structure in the field distribution is important for the delay incurred 
by secondary photons produced by the decay of photo-produced neutral pions 
(Waxman \& Coppi, 1996): in the impulsive scenario, the delay of secondary  
photons will mirror the  delay incurred by the UHECRs at their production site 
with respect to the primary photons associated with the explosive event.

If the IGM magnetic field distribution is strongly structured on scales
of $\gesim 10$ Mpc, the observational upper limit for the typical 
field strength, $B_{\rm r} \leq 10^{-9}$ G, does not apply. Random 
fields on Mpc scales, concentrated in cluster or superclusters, could be as 
strong as $10^{-7}-10^{-6}$ G. In that case, a totally different model of
UHECR propagation should be considered, where particles at lower energy are
trapped within clusters in a manner reminiscent of the `Leaky Box' models often
used in the description of the propagation and confinement of Galactic cosmic rays
(Rachen, 1998 {\em private communication}). 
Such a description would extend to particles with energy 
$E \lesim 10^{20} \: Z \: (B_{\rm r}/0.1 \; \mu {\rm G}) \: (L/{\rm Mpc})$ eV,
where $L$ is the size of the cluster.

\section{UHECR energy losses}

UHECRs with energies above $10^{18.5}$ eV lose energy primarily 
through reactions with photons of the cosmic microwave background 
(e.g. Berezinskii \etal, 1990 and references therein). 
Because of the strong energy dependence of scattering on IGM magnetic 
fields, ${\cal D}_{0} \propto E^{-2}$, this energy evolution has to be 
considered in detail when describing the intergalactic propagation of UHECRs.
We parameterise the mean energy loss per unit path length by the energy
loss length $\ell(E)$ defined by
\begin{equation}
	\ell(E) = 
	\left|\frac{1}{E} \: \frac{{\rm d} E}{{\rm d} s} \right|^{-1}
	\; .
\end{equation}
At the lowest energies under consideration in the simulations we will
describe in the next Section, $10^{17} \le E \le 10^{18}$ eV, the dominant 
loss mechanism is (adiabatic) expansion losses in the Hubble flow. The loss length 
equals the Hubble length,
\begin{equation}
	\ell_{\rm H} \simeq \frac{c}{H_{0}} \simeq 3 \: h^{-1} \; {\rm Gpc} \; .
\end{equation}
For protons in the energy range $10^{18.5} < E < 10^{19.5}$ eV, 
the dominant loss
mechanism is photon pair production: 
$p + \gamma \rightarrow p + e^{+} + e^{-}$
(Blumenthal, 1970). The energy loss length $\ell_{\rm p}(E)$ 
for this process reaches its minimum value in the energy range 
$10^{18.5} < E < 10^{19.5}$ eV 
(see the figure below and also fig. 3 of Yoshida \& Teshima, 1993):
\begin{equation}
	\ell_{\rm p} (E)  
	\simeq 2 \; {\rm Gpc} \; .
\end{equation}
At higher energies, losses are dominated by photo-pion production:
$p + \gamma \rightarrow p + \pi$'s.
As more and more CMWB photons in the Wien tail of the blackbody spectrum
satisfy the threshold condition for pion production at increasing proton 
energies, the associated loss length $\ell_{\pi}(E)$ decreases exponentially 
with increasing energy. In the energy range $10^{19.5} < E < 10^{20.5}$ eV 
it scales as (see the discussion below)
\begin{equation}
\label{pilength}
	\ell_{\pi}(E) \simeq 4.8  \: 
	\left( \frac{E}{E_{\rm th}}\right)^{2} \;
	e^{E_{\rm th}/E}  \; {\rm Mpc}\; ,
\end{equation} 
where $E_{\rm th} \simeq  (m_{\pi} \mpr c^{4})/(2 k_{\rm b} T_{\rm CMWB}) 
\simeq 3 \times 10^{20} \; {\rm eV}$ is the threshold energy. 	
For $E \ge 10^{20.5}$ eV pion production on the CMWB saturates at a 
constant loss length, as essentially all CMWB photons exceed the threshold
in the proton rest frame:
\begin{equation}
	\ell_{\pi}(E) \simeq 10 \; {\rm Mpc} \; .
\end{equation}
The energy loss length in the local universe ($z = 0$) resulting from
these combined processes is shown in Figure 1.

\begin{figure}%[hbtp]
\centerline{\psfig{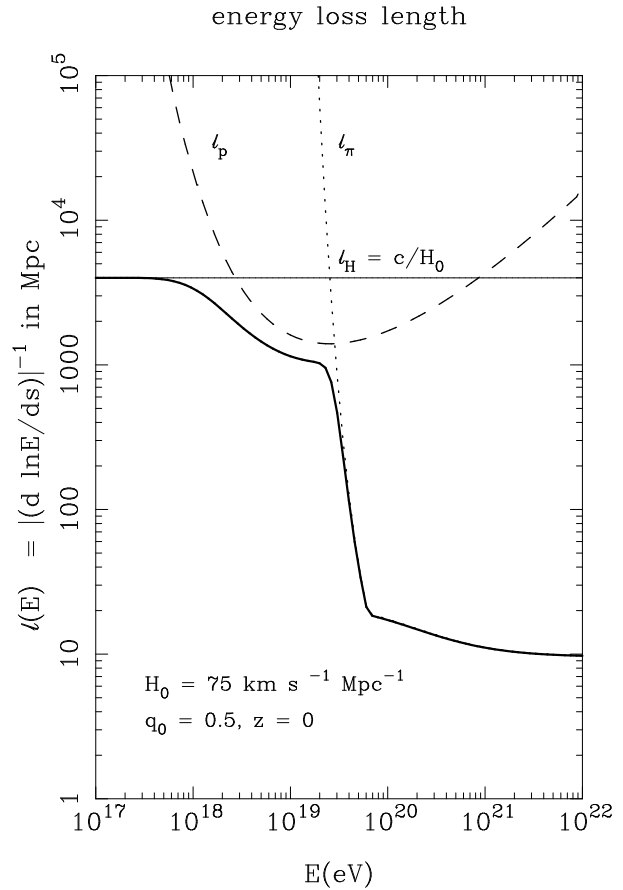}}
\caption{ The loss length $\ell(E)$ (thick solid curve) as a function
of energy. It is the result of losses due to pion production ($\ell_{\pi}$, 
dotted curve),
pair losses ($\ell_{\rm p}$, dashed curve) and expansion losses in the 
Hubble flow ($\ell_{\rm H} = c/H_{0}$, horizontal solid line)
in a flat Universe ($q_{0} = \mbox{$\frac{1}{2}$}$).}
\end{figure}
 
In order to describe the UHECR spectrum received at Earth,
one has to calculate the modification of the spectrum at
the source due to pion-production on the CMWB.
A number of pioneering studies have been done in the past
(e.g. Berezinskii, Grigor'eva \& Zatsepin, 1975; Hill \& Schramm, 1985;
Berezinskii \& Grigor'eva, 1988; Rachen \& Biermann, 1993; 
Yoshida \& Teshima, 1994 and Aharonian \& Cronin, 1994).

Much of the discussion has focussed on the approximation to be
used to describe the spectrum near the Greisen-Zatsepin-Kuz'min
(GZK) cut-off in the spectrum (Greisen, 1966; Zatsepin \& Kuz'min, 1966), 
which results from the exponential dependence of the 
photo-pion production loss length (Eqn. \ref{pilength}) .
This cut-off occurs at an energy $E_{\rm c} \sim 10^{19.5}$ eV.
Particles injected above the cut-off are expected to `pile up' close to
$E_{\rm c}$, forming a bump in the spectrum.
The height and width of this bump is determined by the details of
the pion production process.

Early work (e.g. Berezinskii \& Grigor'eva, 1988) assumed that
the effect of pion losses could be described by a simple
continuous energy loss approximation, i.e. ${\rm d} E/{\rm d} s 
= - E/\lE$.
It was pointed out by Hill \& Schramm (1985) 
that there is an  intrinsic spread in energy which is not described
by this simple approximation. This spread results from two effects:
\begin{enumerate}
\item	The  kinematics of the photo-pion production, which leads
	to a spread in the energy of particles in the observer's
	frame due to the spread of pion production angles in the
	center-of-momentum frame (CMF).
	
\item	The Poisson noise in the number of photons encountered for 
	a given path length $s$.
\end{enumerate}
These effects modify the result of the continuous loss approximation, 
spreading the bump over a larger energy range and making it less prominent.
In the next section, we will consider a simple model for the evolution
of the UHECR spectrum  in transit from source to observer due to these effects.

\subsection{Kinematics of photo-pion production}

Because of its importance in determining the GZK cut-off energy, and the
`survival probability' of UHECRs above the cut-off, we consider photo-pion
production in somewhat more detail. A more complete account can be found in
Berezinskii \etal (1975) or in Mannheim \& Biermann (1989).
All expressions are for protons ($Z = A = 1$), but can be readily generalised
to other nuclei. We use units with $c = 1$ in this Section.

The interaction between a proton with energy $\Ep$ and a photon with energy
$\epsilon$ in the observer's frame is most conveniently described in terms
of the invariant total energy $\Et$ in the center of momentum frame (CMF),
which moves with Lorentz factor $\gamma_{\rm CMF} = (\Ep + \epsilon)/\Et
\simeq \Ep/\Et \gg 1$, and the photon energy $\eprf$ in the proton rest 
frame (PRF). They are related by
\begin{equation}
	\Et =   \sqrt{\mpr^{2} + 2 \mpr \eprf} 
	\;  \; , \; \; 
	\eprf =  \gp \epsilon \: \left( 1 - \bp \cos{\theta} \right)
	\; .
\end{equation}
Here $\gp = 1/\sqrt{1 - \bp^{2}}$ is the Lorentz factor of the
proton, and $\theta$ the angle between the photon propagation
direction and the proton momentum in the lab frame. 
A head-on collision between proton and photon corresponds to $\theta = \pi$.

Energy and momentum conservation in the CMF, followed by a Lorentz
transformation back to the observers frame, yield the following
expression for the final proton energy:
\begin{equation}
\label{Echange}
	\Ep^{\rm f} \simeq 	
	\Ep \: 
	\left[ 1 - K_{\rm p} + \tilde{K} \: \cos{\sigma_{\rm if}} \: \right]
	\; .
\end{equation}
Here $\sigma_{\rm if}$ is the angle between the 
momentum vector of the incoming and final proton in the CMF.
We have put $\beta_{\rm CMF} \approx 1$, which is an excellent approximation
at these energies, and defined a {\em mean elasticity}
\begin{equation}
\label{inelast}
	K_{\rm p} \equiv \frac{\Et^{2} + \mpi^{2} - \mpr^{2}}{2 \Et^{2}} \; .
\end{equation}
The quantity $\tilde{K}$ is a measure of the spread of final energies around the 
mean,
\begin{equation}
\label{Edisp}
	\tilde{K} \equiv 
	\frac{\displaystyle \sqrt{(\Et^{2} - \mpl^{2})(\Et^{2} - \mmi^{2})}}
	{2 \Et^{2}}
	\; ,
\end{equation}
with $\mpm \equiv \mpr \pm \mpi$.
If one assumes that the CMF production angle $\sigma_{\rm if}$ is 
distributed isotropically, the {\em mean} energy loss per interaction
in the observer's frame equals $\Delta \Ep = - K_{\rm p} \Ep$ and the
energy dispersion is $\Delta E_{\rm rms} \simeq (\tilde{K}/\sqrt{3}) \: \Ep$.

The spread in the final proton energy in the observer's frame increases
strongly with CMF energy.
Near reaction threshold, which corresponds to zero momentum for the final proton 
and the production of a single pion at rest so that $\Et \simeq \mpr + \mpi$, 
one has $K_{\rm p} \simeq 0.13$ and $\tilde{K}\approx 0$, and the
spread in the final proton energies in the  observer's frame is small. 
Much above threshold, where $\Et \gg \mpl$ and
$K_{\rm p} \approx \tilde{K} \approx \frac{1}{2}$, one has
$\Ep^{\rm f} \approx \frac{1}{2} \: \Ep \:(1 + \cos{\sigma_{\rm if}})$,
so the spread in final energies is large.

The interaction rate for photo-pion production in the observer's frame, 
${\cal R}_{\rm p \gamma}$, can be expressed in terms of the reaction 
cross-section $\sigma_{0}^{\rm p \gamma}(\epsilon_{0})$ in the PRF by 
using the Lorentz-invariance of the photon occupation number in phase space 
and the relativistic aberration formulas which, in the limit 
$\gamma_{\rm p} \gg 1$, lead to strong relativistic beaming of the 
CMWB photons in the PRF, so that all photons have $\cos{\theta_{0}} \simeq - 1$. 
The relevant expressions can be found in Blumenthal (1970).

This leads to an expression for the reaction rate 
derived by Berezinskii {\em et al.} (1975), 
but with corrected integration limits corresponding to the 
kinematically allowed values:
\begin{equation}
\label{totrate}
	{\cal R}_{\rm p \gamma} = 
	\int_{\epsilon_{\rm th}/ 2 \gp}^{\infty} {\rm d} \epsilon \: 
	\frac{n(\epsilon)}{\epsilon^{2}} \: \left( \frac{1}{2 \gp^{2}} \:
	\int_{\epsilon_{\rm th}}^{2 \gp \epsilon} {\rm d} \epsilon_{0} \: 
	 \:\sigma_{0}^{\rm p \gamma}(\epsilon_{0}) \:
	\epsilon_{0} \right) \; .		
\end{equation}
Here $\epsilon_{\rm th} = \mpi(1 + \mpi/2\mpr) \simeq 145$ MeV is the photon
threshold energy in the PRF.
Berezinskii {\em et al.} (1975) incorrectly replace the upper limit on 
the integration over $\epsilon_{0}$ for fixed $\epsilon$ by infinity.
For a thermal (Planckian) photon distribution in the observer's frame with 
temperature $T$ one has, reinstating $c$:
\begin{equation}
	\frac{n(\epsilon)}{\epsilon^{2}} = \frac{1}{\pi^{2}(\hbar c)^{3}} \:
	\left[ e^{\epsilon/k_{\rm b}T} - 1 \: \right]^{-1} \; .
\end{equation}

The PRF pion production cross-section $\sigma_{0}^{\rm p \gamma}(\epsilon_{0})$
near threshold is dominated by a resonance in the
$p + \gamma \longrightarrow p + \pi^{0}$ 
reaction near $s = \Et^{2} \simeq 1.6 \; {\rm GeV}^{2} \equiv s_{\rm m}$
with $\sigma^{\rm p \gamma}(s_{\rm m}) \simeq 0.5 \: {\rm mb}$.
Therefore, not too far from threshold, we can approximate the
cross-section by
\begin{equation}
	\sigma^{\rm p \gamma}(s) \simeq 
	\Delta s \; \sigma^{\rm p \gamma}(s_{\rm m}) \: 
	\delta(s - s_{\rm m})
	\; 
\end{equation}
in the integral over $\epsilon_{0}$ or $s$. Here 
$\Delta s \simeq \mpr \epsilon_{\rm th} \simeq 0.15 \; {\rm GeV}^{2}$ is
the width of the resonance.
The value of $s_{\rm m}$ corresponds to $\epsilon_{0} \simeq 0.3$ GeV 
$\approx 2 \epsilon_{\rm th}$.

Using the above
approximation for the cross-section, the reaction rate for $\Ep \ll \Etr$ 
typically equals
\begin{equation}
	\cR_{\rm p \gamma}(\Ep) \simeq
	\frac{4}{\pi^{2}}  \: 
	\left( \frac{ k_{\rm b} T}{\hbar c} \right)^{3} c \sigma_{\rm m}
	\: \left(\frac{\Etr}{\Ep} \right)^{2} \:
	e^{-E_{\rm th}/E_{\rm p}} \; ,
\end{equation}
with $\sigma_{\rm m} = \sigma^{\rm p \gamma}(s_{\rm m})$,
and $\Etr = (m_{\pi} \mpr c^{4})/(2 k_{\rm b} T)$.
Putting $\Etr^{2} = s_{\rm m}$ in Eqns. (\ref{inelast}) and (\ref{Edisp})
yields $K_{\rm p} \simeq 0.2$ and $\tilde{K} \simeq 0.165$ in this limit. 

At energies $\Ep \gg \Etr$, $\Et \gg \mpr + \mpi$, the cross-section 
summed over all reactions becomes approximately constant, 
$\sigma^{\rm \pi \gamma}(s) \simeq 0.16 \: {\rm mb} \equiv \sigma_{0}$.
The reaction rate $\cR^{\rm p \gamma}(\Ep)$ then approaches 
the constant rate
\begin{equation}
	\cR_{\rm p \gamma}(E_{\rm p} \gg E_{\rm th}) \simeq 0.244 \:
	\left( \frac{ k_{\rm b} T}{\hbar c} \right)^{3}
	\: c \sigma_{0} \; ,
\end{equation}
while the mean fractional energy change and energy dispersion 
satisfy $K_{\rm p} \simeq \tilde{K} \simeq 0.5$. 
The energy loss length $\ell(E)$ follows from these quantities as
\begin{equation}
\label{lossdef}
	\ell(E) = c/ \left({\cal R}_{{\rm p} \gamma}(E) \: K_{\rm p} \right) \; .
\end{equation}

\subsection{Effects of Poisson noise}

The {\em mean} number of pion-producing photons encountered by 
UHECR particles in a path length $\Delta s = c \Delta t$ equals
\begin{equation}
	\eave{n_{\rm ph}}(\Delta s) \simeq  
	\frac{ \cR_{\rm p \gamma} \: \Delta s}{c}
	= \frac{\Delta s}{K_{\rm p} \ell(E)} \; .
\end{equation}
The {\em actual} number of photons encountered by
an individual particle is subject to Poisson statistics.
This means that the variance in the number of encounters equals
$\chi_{\rm n} = \eave{n_{\rm ph}}^{1/2}$,
while the probability of a particle not encountering {\em any} 
photons decays as $P(n_{\rm ph} = 0) = e^{- \eave{n_{\rm ph}}}$.
The effect of Poisson statistics is especially important for protons 
at high energies, $\Ep \ge 10^{18.5}$ eV, where $K_{\rm p} \ell(E) \simeq 20$ Mpc. 
For protons originating from sources closer than $D \sim 20-50$ Mpc, the effect 
of Poisson statistics on the energy loss should be clearly visible in the arrival 
spectrum. There is a significant fraction of particles which have 
interacted with no (or only a few) photons, leading to a high-energy tail in 
the spectrum which more or less reflects the original source spectrum,
because the loss length $\ell(E)$ is roughly independent of energy, 
but at an amplitude reduced by a factor $\sim e^{- \eave{n_{\rm ph}}(D)}$.

\section{Simulations}

We have constructed a simulation code for the propagation of protons in
intergalactic space, taking into account the scattering on IGM magnetic 
fields and the energy losses suffered as a result of interactions with 
the CMWB.

The propagation of the UHECRs over a path length $\Delta s$ is
calculated with the simple explicit scheme
\begin{equation}
	\bm{x}(s + \Delta s) = \bm{x}(s) + \hatn(s) \: \Delta s \; .
\end{equation}
The unit vector $\hatn$ along the direction of propagation undergoes diffusion,
which is  described by the the explicit numerical integration of a
stochastic differential equation (SDE). The stochastic change in the direction 
of flight is modelled by 
\begin{equation} 
	\Delta \hatn_{\rm st} = \sqrt{2 {\cal D}_{0} \: \Delta s} \: 
	\left( \xi_{1} \evec{1} + \xi_{2} \evec{2} \right) \; .
\end{equation}
Here $\xi_{1}$ and $\xi_{2}$ are two independent unit Wiener processes, 
drawn at each step from a Gaussian distribution with unit dispersion so 
that $\left< \xi_{1,2} \right> = 0$, $\left< \xi_{1,2}^{2} \right> = 1$. 
The two (arbitrary but mutually orthogonal) unit vectors $\evec{1,2}$ 
lie in a plane perpendicular to $\hatn$ so that 
$\Delta \hatn_{\rm st} \bdot \hatn = 0$.
Note that $\hatn \equiv (n_{1} \: , \: n_{2} \: , \: n_{3})$ and 
$\evec{1,2}$ are the eigenvectors (with eigenvalues $0$ and $1$ respectively)
of the diffusion tensor ${\cal D}_{ij} = {\cal D}_{0} \: P_{ij}$, with
$P_{ij} \equiv \left( \delta_{ij} - n_{i} n_{j} \right)$, 
which motivates this approach. 
This prescription is equivalent to the stochastic
term in Eqn. (\ref{Itosde}) of Appendix B, with the correspondence
\begin{equation}
	\sqrt{2 {\cal D}_{0}} \: \bm{P}(\hatn) \bdot \Delta \bm{W}
	\longrightarrow \sqrt{2 {\cal D}_{0} \: \Delta s} \: 
	\left( \xi_{1} \evec{1} + \xi_{2} \evec{2} \right) \; .
\end{equation}		 
The new propagation direction then is calculated as
\begin{equation}
	\hatn(s + \Delta s) = \sqrt{1 - \left| \Delta \hatn_{\rm st} \right|^{2}} 
	\:\hatn(s) +  \Delta \hatn_{\rm st} \; .
\end{equation} 
This preserves the norm $\hatn \bdot \hatn = 1$ 
and is an excellent approximation for the diffusion 
process provided $| \Delta \hatn_{\rm st} | \ll 1$.
When many independent realizations of these SDEs are simulated, the 
resulting distribution of unit vectors corresponds to a solution of 
the diffusion equation for $\hatn$.
Details of similar astrophysical applications of SDEs can be found in 
Achterberg \& Kr\"ulls (1992).

The change in UHECR energy is calculated as the combined effect of
expansion losses in the the Hubble flow, pair production and photo-pion 
production losses. 
Pair production is treated using the continuous loss approximation, 
$({\rm d} E/{\rm d} s)_{\rm p} = - E/\ell_{\rm p}$, which is 
an excellent approximation for this process.
We use the interpolation formula of Stepney \& Guilbert (1983) for the
cross-section as a function of proton energy, 
and the numerical fit of Chodorowski, Zdziarski \& Sikora 
(1992)  for the inelasticity of the reaction. The integration over the 
CMWB photon  distribution in the proton rest frame proceeds along the lines 
elaborated in Section 3. The pair production loss length $\ell_{\rm p}(E)$ so 
obtained (see Figure 1) closely follows the result of Blumenthal (1970). 

Pion production must be treated in more detail. The number of pion-producing 
photons encountered in a path-length increment $\Delta s$ is drawn from a 
Poisson distribution with mean $\eave{n_{\rm ph}} = \Delta s/\lpg$.
The interaction length $\lpg \equiv c/{\cal R}_{\rm p \gamma}$ is 
calculated according to (with threshold energy $\Etr \sim 3 \times 10^{20}$ eV):
\begin{equation}
	\lpg(\rm Mpc) =  \mbox{$\left\{ \begin{array}{ll} 
	0.9 \: {\displaystyle \left( \frac{E}{\Etr} \right)^{2} \: 
	e^{E_{\rm th}/E}}
	& \mbox{($E \le 0.2 \Etr$),} \\
	& \\ 
	& \\
	4.8 \;  & \mbox{($E > 0.2 \Etr$).} 
	\end{array} \right. $}
\end{equation}
Here we have written $E$ rather than $\Ep$ for the proton energy.
The energy loss per interaction is calculated using a simple
interpolation formula for the mean inelasticity: 
\begin{equation}
	K_{\rm p} =   
	0.2 \: \left(\frac{\Etr + 2.5 E}{\Etr + E} \right) \; .
\end{equation}
The pion production angle in the center of momentum frame is drawn from a 
uniform distribution $-1 \le \cos{\sigma_{\rm if}} \le 1$. 
The energy change is calculated according to Eqn. (\ref{Echange}), 
with $\tilde{K}$ calculated by re-expressing $\Et^{2}$ in terms
of $K_{\rm p}$ using the exact kinematic relations for $K_{\rm p}$ and 
$\tilde{K}$:
\begin{equation}
	\tilde{K} \equiv
	\sqrt{\displaystyle 
	\left(K_{\rm p} + K_{+} \right) 
	\left(K_{\rm p} - K_{-} \right)} \; ,
\end{equation} 
with $K_{\pm} \equiv \mpi/(\mpr \mp \mpi)$.
This allows the use of a single interpolation for $K_{\rm p}(E)$ while 
preserving the correct behaviour of $\tilde{K}$ at both low and high energies.
This is illustrated in Figure 2.

\begin{figure}%[hbtp]
\centerline{\psfig{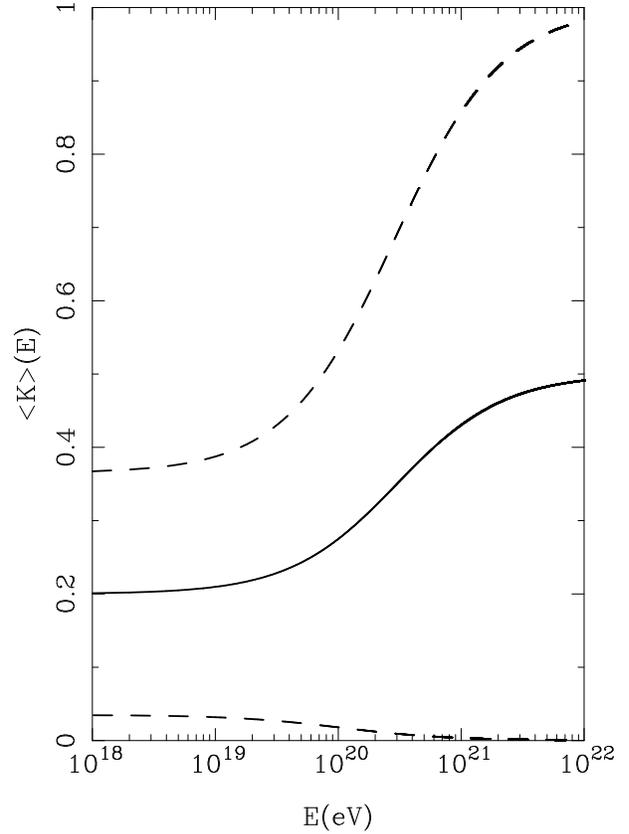}}
\caption{ The mean inelasticity $\left< K \right> \equiv K_{\rm p}$
(solid line) and the upper and lower
limit (dashed lines), corresponding to \mbox{$\left< K \right> \pm \tilde{K}$}, 
due to the variance in production angle $\sigma_{\rm if}$ in the PRF.
At given energy, the inelasticity is uniformly distributed between the
dashed lines.}
\end{figure}

Finally, at large source distances cosmological effects are taken into account.
Due to the higher CMWB photon density and temperature at a distance
corresponding to a redshift $z$, 
$n_{\rm ph} \propto (1+z)^{3}$ and $T_{\rm CMWB} \propto (1 + z)$,
the energy loss length due to interaction with CMBW photons at redshift $z$ 
can be obtained from the value in the local universe
($z = 0$) by using the scaling law
\begin{equation}
	\ell\left( E,z \right) = \frac{\ell((1+z)E,z=0)}{(1 + z)^{3}} \; .		
\end{equation}
This is a simple consequence of the fact that the loss length is inversely 
proportional to the photon number density and depends on energy $E$ and the 
lab frame threshold energy 
$E_{\rm th} \propto T_{\rm CMWB}^{-1} \propto (1+z)^{-1}$ 
of the interaction through the ratio $E/E_{\rm th} \propto 1 + z$.

The random magnetic field strength in the general intergalactic medium
scales as $B_{\rm r} \propto (1 + z)^{2}$ as a result of flux freezing
in an expanding universe (e.g. Subramanian \& Barrow, 1998). The universal
expansion stretches the correlation length so that $\lc(z) =
\lc(z=0)/(1+z)$. As a result of these effects the magnitude of the 
flight direction diffusion rate scales as 
${\cal D}_{0} \propto B_{\rm r}^{2} \lc \propto (1 + z)^{3}$.
In our simulations we assume a flat, matter-dominated universe in which the 
Hubble length scales as $\ell_{\rm H} = (c/H_{0}) \times (1+z)^{-3/2}$.

\subsection{Simulation results}

To illustrate the effects of UHECR scattering most clearly, 
we consider the case of mono-energetic injection first. 
Particles are injected with an energy $E_{\rm inj} = 10^{21}$ eV.
Arrival spectra, delay times and angular distributions are calculated
for different source distances. Figure 3 shows the arrival spectra.

\begin{figure}%[hbtp]
\centerline{\psfig{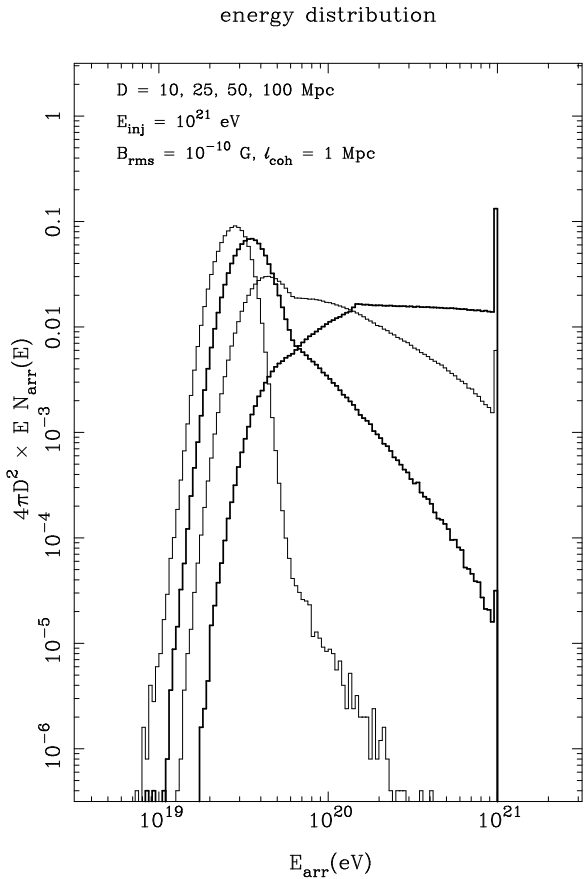}}
\caption{ Arrival spectra at Earth for mono-energetic injection at 
an energy of $10^{21}$ {\rm eV} at the source,
for four source distances $D$.
The spectra have been multiplied by an 
arbitrary factor $\propto D^{2}$.
For increasing distance, the maximum of the spectrum
shifts to lower energy. For 10 {\rm Mpc} there is still a significant
fraction of particles at the injection energy, giving rise to
a distinct spike. This spike has almost completely vanished
for $D = 50$ {\rm Mpc}.
The spectrum above $10^{19.5}$ {\rm eV}, where the loss length
$\ell(E)$ decreases by almost two orders of magnitude, decreases 
rapidly for larger distances.}
\end{figure}

From this calculation it is obvious that the effect of Poisson statistics
remains strong up to distances of about $50$ Mpc: a tail of high-energy particles 
up to the injection energy remains. At larger distances, the probability of not 
encountering any CMBW photons becomes vanishingly small. Energy losses proceed much 
more slowly once particles reach an energy $E < 10^{19.5}$ eV, as the loss length 
becomes larger than $1$ Gpc. Therefore particles tend to accumulate around
$10^{19.5}$ eV, drifting slowly to lower energies.

Figure 4 shows the distribution of the delay time with respect to
simultaneously produced photons, for a number of different source distances.
At small distances they are centered around a value close to the 
predicted value of $t_{\rm del}$ for the injection energy of $10^{21}$ eV.
For those distances larger than 25 Mpc, the delay distribution has its
maximum close to the value of $t_{\rm del}$ associated with the typical
arrival energy, $E_{\rm arr} \simeq 10^{19.5}$ eV.

Figure 5 shows the cumulative angular distribution of the UHECRs
for the arrival angle $\alpha'$ with respect to the line of sight 
to the source. The deflection process considered here will lead 
{\em for constant particle energy} to an angular distribution of the form
${\rm d} N/{\rm d} \Omega \propto {\rm exp}(- \alpha'^{2}/\alpha_{\rm rms}^{2})$,
with ${\rm d} \Omega = 2 \pi \sin{\alpha'} \: {\rm d} \alpha'$.
For small angles, $\alpha'$ and $\alpha_{\rm rms} \ll 1$, this 
distribution scales as
\begin{equation}
\label{modeldistr} 
	\frac{{\rm d}N}{{\rm d}\ln \alpha'} \propto 
	\alpha'^{2} \: 
	e^{-\alpha'^{2}/\alpha_{\rm rms}^{2}} \; , 
\end{equation}
which has a maximum at $\alpha' = \alpha_{\rm rms}$. The 
simulated distributions shown in Figure 5, calculated with energy losses taken 
into account, closely resemble model distribution (\ref{modeldistr}).

\begin{figure}%[hbtp]
\centerline{\psfig{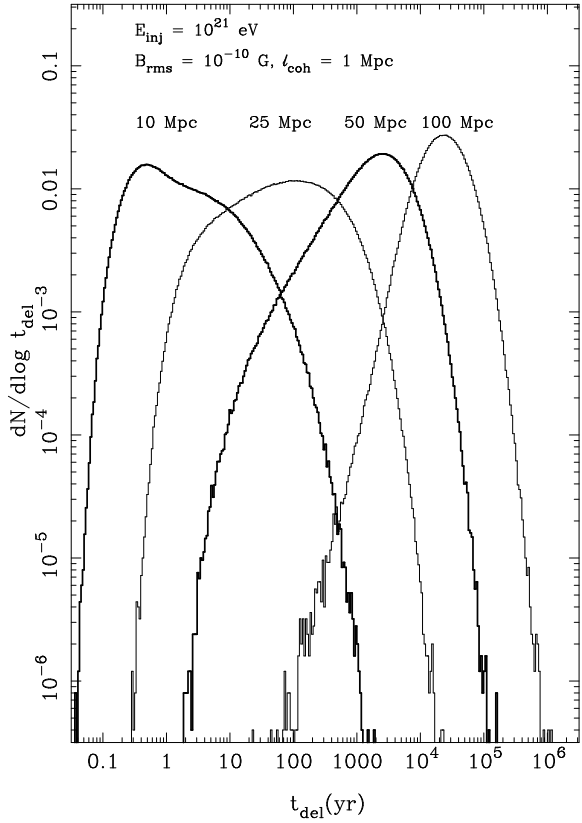}}
\caption{ Delay time distribution at Earth for 
particles mono-energetically injected at $E = 10^{21}$ {\rm eV} for four 
source distances ranging from $10$ to $100$ {\rm Mpc}. 
The distribution for the smallest distance is the
left-most curve, for the largest distance the right-most curve.
The maximum in the $10$ {\rm Mpc} curve coincides with the
theoretical result of Eqn. \ref{del value}
for that distance at the injection energy,
$t_{\rm del}(10^{21} \; {\rm eV}) \simeq 0.3$ year. 
The maximum for the $100$ {\rm Mpc} curve 
coincides roughly with the theoretical result for 
that distance and the energy where most particles reside (see figure 3):
$E \sim 2 \times10^{19}$ {\rm eV}, yielding $t_{\rm del} \simeq 10^{5}$ {\rm year}}.
\end{figure}

\begin{figure}%[hbtp]
\centerline{\psfig{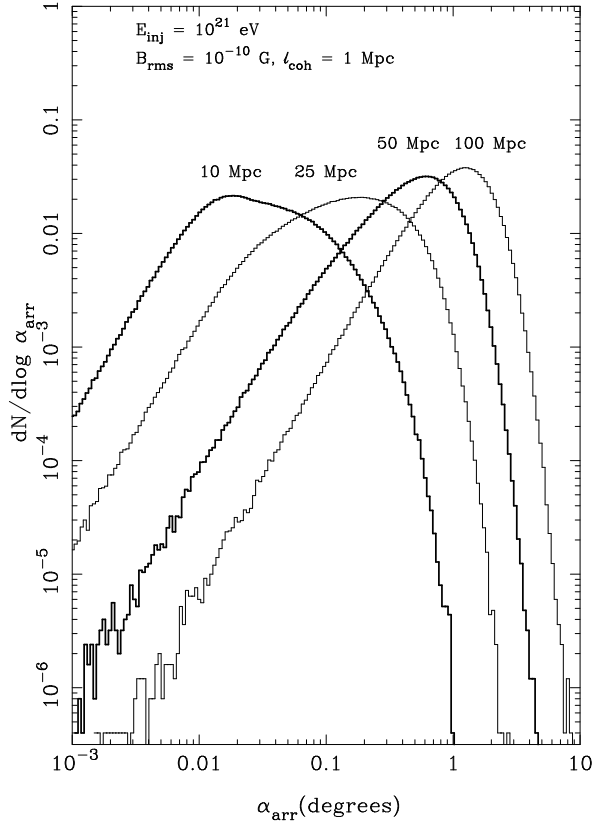}}
\caption{ Angular distribution at Earth of the arrival angle
$\alpha_{\rm arr} \equiv \alpha'$ with respect to the line-of-sight
to the source, for particles monochromatically injected at $E = 10^{21}$ eV 
for distances ranging from 10 to 100 {\rm Mpc}.
The maximum of these curves is roughly at $\alpha_{\rm rms}$
(Eqn. \ref{alpha rms}).
For a distance of $10$ {\rm Mpc} and $E \sim E_{\rm inj} = 10^{21} \; {\rm eV}$,
theory predicts $\alpha_{\rm rms} \simeq 0.012$ {\rm degrees}.
For a distance of $100$ {\rm Mpc} and an energy of $E \sim 2 \times 10^{19}$ {\rm eV},
one expects $\alpha_{\rm rms} \simeq 2$ {\rm degrees}.}
\end{figure}

Next we consider the case of power-law injection, where particles are produced at
the source according to
\begin{equation}
	{\rm d} N_{\rm inj}(E) = n_{\rm inj}(E) \: {\rm d}E =
	\kappa \: E^{-s} \: {\rm d} E \; .
\end{equation}
Figure 6 shows the arrival spectrum at Earth as expected for a steady source.
The injection spectrum was $n(E) \propto E^{-2}$ between $10^{17}$ and $10^{21}$
eV. The delay incurred in transit has a negligible influence
on the shape of the spectrum as long as $ct_{\rm del} \ll D$. 
Therefore the arrival spectra calculated for $B_{\rm rms} = 10^{-11}$ G and
$\lc = 1$ Mpc are representative for all simulations we performed at energies
well above $E_{\ast} \approx 10^{18} Z B_{-9} \ell_{0}$ eV.
The spectra have been multiplied with a factor $\propto D^{2} \: E^{2}$ 
so that the injection spectrum in all cases corresponds to a horizontal line.
For energies below $10^{19}$ eV, the effect of losses is negligible at
these source distances. Particles do accumulate at the GZK cut-off energy,
but the height of the `bump' is not very dramatic due to the natural spread in arrival
energies, $\Delta E/E \simeq {\cal O}(1)$ for a given injection energy due to the
variance in the photo-pion production kinematics.  
The high-energy tail above the GZK cut-off is very pronounced at $D = 10$ and 
at $D = 25$ Mpc, still quite visible in the
$50$ Mpc curve, but almost absent at a source distance  $D = 100$ Mpc.
Our results are qualitatively similar to those obtained by 
Yoshida \& Teshima (1993) in the corresponding distance range.

\begin{figure}%[hbtp]
\centerline{\psfig{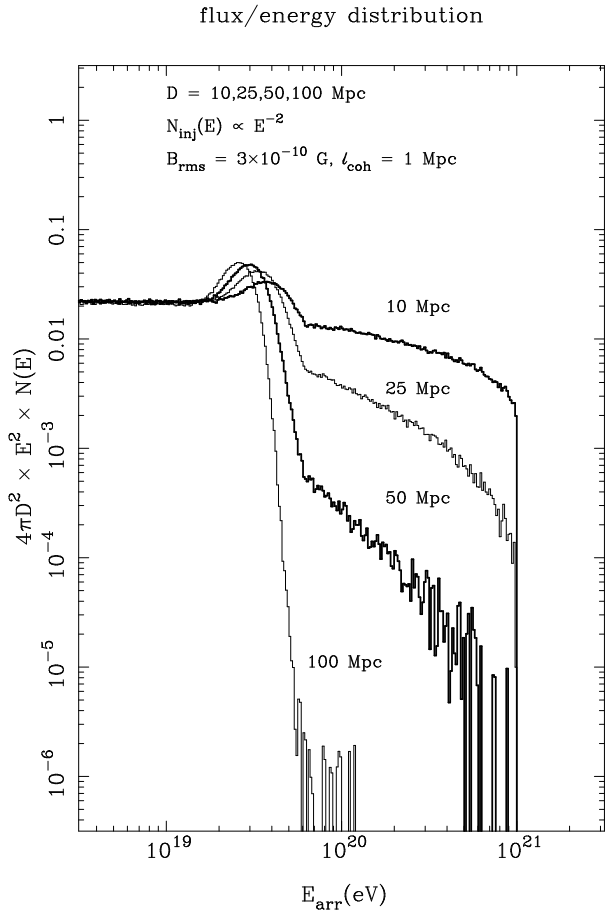}}
\caption{ Energy distribution $n(E) \times E^{2}$,
received at Earth for a continuously radiating source. 
Particles are injected at the source with $n_{\rm inj}(E) \propto E^{-2}$ 
between
$10^{17} \le E \le 10^{21}$ {\rm eV}. Spectra are calculated for 
a source distance $D$ equal to 
$10$, $25$, $50$ and $100$ {\rm Mpc}, and multiplied by an arbitrary factor
$\propto D^{2}$.  Only the portion of the 
spectrum between $10^{18.5}$ and $10^{21.5}$ {\rm eV} is shown. 
This figure should be compared with figure 3 for monochromatic injection.
There is a clear `bump' near the GZK cut-off at $E \sim 10^{19.5}$ {\rm eV}.
Below the GZK cut-off, particles lose energy very slowly
($\ell(E) \ge 1$ {\rm Gpc}), and the spectrum is almost independent
of source distance.}
\end{figure}

Figure 7 shows the delay time $t_{\rm del}$ and the dispersion
$\Delta t_{\rm del}$ (shown with error bars as 
$t_{\rm del} \pm \Delta t_{\rm del}$) for particles originating at 
$D = 25$ Mpc as a function of the {\em arrival} energy $E_{\rm arr}$ at Earth.
The delay time has been scaled by a factor 
$\propto (E_{\rm arr}/10^{20} \; {\rm eV})^{2}$
to scale away the dominant energy dependence predicted by Eqn. (\ref{delay time}).
Particles well below the GZK cut-off ($E < 10^{19}$ eV) have lost little
energy, and therefore fall almost on the horizontal line predicted
by Eqn. (24) for $Z = 1$ (protons):
\begin{equation}
	t_{\rm del}(E,D) \: \ell_{0}^{-1} \: (E_{20}/B_{-9} D_2)^{2}
	\simeq 0.31 \; {\rm Myr} \; . 
\end{equation}
At the lowest energies considered here, $E \simeq 10^{17}$ eV, there is
an increase in the delay time because the small deflection angle approximation
starts to break down: $\alpha_{\rm rms} \simeq 20^{o}$ at these energies.
Well above the GZK cut-off the delay time is roughly half the expected value based on 
the arrival energy due to the fact that the mean energy in transit is larger
that the arrival energy.
For energies near the injection energy, $E \approx 10^{21.5}$ eV, the delay
time once again approaches the theoretical value as these are particles that
happen not to have interacted significantly with CMWB photons, and have
therefore propagated with an almost constant energy.

\begin{figure}%[hbtp]
\centerline{\psfig{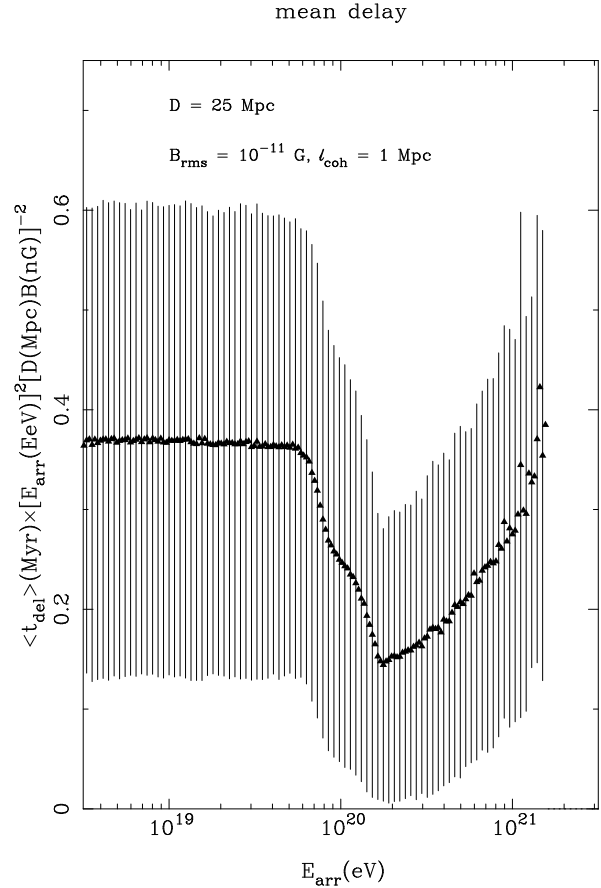}}
\caption{ Delay time as a function of arrival energy. The dispersion
in the delay time is shown as error bars. The delay has been scaled such
that propagation at {\em constant} energy $E$ would result in a horizontal
line at $t_{\rm del} \times E^{2}_{\rm EeV}/(D_{\rm Mpc} B_{\rm nG})^{2} \simeq 0.31$
{\rm Myr}, cf. Eqn. 24. One sees that for the particles near
$E_{\rm max} = 10^{21.5}$ {\rm eV}, which happen to have lost little energy, 
and for the particles in the range $E < 10^{19.5}$ {\rm eV} which, 
at this source distance, have lost very little energy,
this theoretical value is approximated. 
For $10^{20} \; {\rm eV} < E < 10^{21}$ {\rm eV} the
influence of energy losses reduces $t_{\rm del}$ with respect to the
value predicted on the basis of the arrival energy $E_{\rm arr}$.}
\end{figure}

Figure 8 shows the dispersion in delay times, $\sigma_{\rm del}$, as a
function of the delay time for the same arrival energy bins as in Figure 7, 
with the dominant energy dependence scaled away.
From this figure one can conclude that, at least in the regime considered here,
the often-used approximation (e.g. Miralda-Escud\'e \& Waxman, 1996)
\begin{equation}
	\sigma_{\rm del}(E,D) \equiv
	\sqrt{ \left< \left( t_{\rm del} - <t_{\rm del}> \right)^{2} \right>}
	\simeq t_{\rm del}(E,D)
\end{equation}
is a fair one. We find roughly $\sigma_{\rm del} \approx 0.6 \left< t_{\rm del} \right>$,
but this relation is approximate, and breaks down when the delays (and deflection angles)
become large (not shown).

\begin{figure}%[hbtp]
\centerline{\psfig{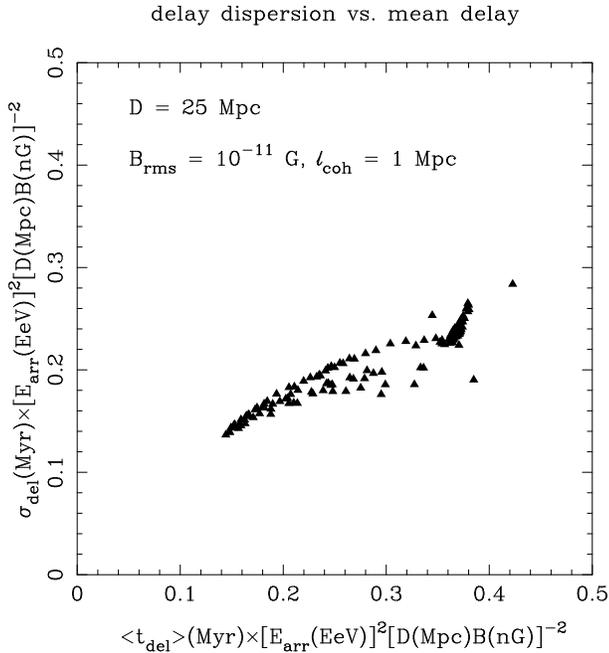}}
\caption{ Delay time dispersion $\sigma_{\rm del}$
as a function of mean delay time $\eave{t_{\rm del}}$
for a source at a distance of 25 {\rm Mpc} in a 0.01 {\rm nG} field.
The dominant energy dependence $\propto E^{-2}$ has been
scaled away. The normalization of the delay time and
delay dispersion is as in figure 7.
At this distance and field strength, where the small-angle approximation 
for the deflection process still applies, one can fit the delay dispersion roughly with
a linear relation of the form $\sigma_{\rm del} \sim 0.6 \eave{t_{\rm del}}$.
}

\end{figure}

Figure 9 shows the angle 
$\alpha' = {\rm cos}^{-1}(\hatn \bdot \bm{\hat{r}})$
with respect to the line of sight to the source as a function of arrival energy.
The dominant behaviour, $\alpha' \propto E^{-1}$, has been scaled away.
If particles did not lose energy, the arrival angle would satisfy (Eqn. 22)
\begin{equation}
	\alpha'
	\times \left( \frac{E_{20}}{B_{-9}} \right) \:
	\left( D_{2} \ell_{0} \right)^{-1/2} \simeq 3.5^{o} \; .
\end{equation}
The actual value (using $E = E_{\rm arr}$)
is somewhat less as a result of the energy losses.

\begin{figure}%[hbtp]
\centerline{\psfig{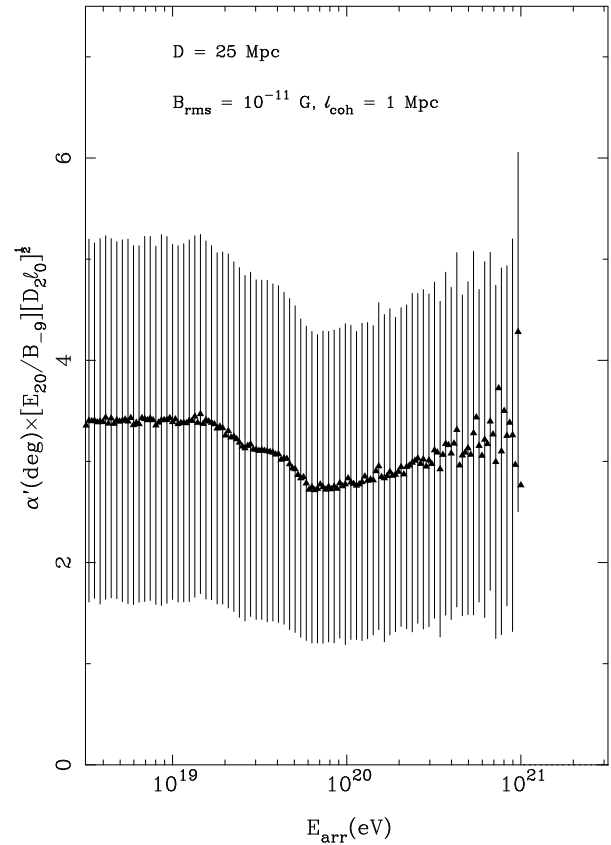}}
\caption{ The arrival angle with respect to the line of sight to the source
as a function of the arrival energy $E_{\rm arr}$.
The dominant parameter dependence $\propto (B \sqrt{D \lc})/E$ has been scaled away
so that particles propagating with constant energy would lie on a horizontal
line $\alpha' \times (E_{20}/B_{-9} \sqrt{D_{2} \ell_{0}}) = 3.5^o$.
Particles have been injected with energies uniformly distributed
over the range $10^{18} < E < 10^{21}$ {\rm eV} 
at a distance of 25 {\rm Mpc}, and propagate through a random field with
$B_{\rm rms} = 10^{-11}$ {\rm G} and $\lc = 1$ {\rm Mpc}.
The dispersion in the arrival angle is shown as bars. Particles with $E < 10^{19}$ {\rm eV}
which lose very little energy for this source distance follow the theoretical
result for constant energy propagation.}
\end{figure}

\section{Bursting sources and the GZK cut-off}

Observations in the northern hemisphere show 
the arrival direction of events above $4 \times 10^{19}$ eV
to be distributed over the whole sky, but with some evidence for clustering
of arrival directions around the supergalactic plane 
(Stanev {\em et al.}, 1995; Hayashida {\em et al.}, 1996).
A similar analysis of the less extensive data set from the southern 
hemisphere (Kewley {\em et al.}, 1996) does not show evidence 
for this clustering.

Due to photo-pion production on the CMWB (Greisen, 1966; Zatsepin \& Kuz'min, 1966)
it is likely that particles with energy above $10^{19.5}$ eV originate 
from a volume with radius $D < D_{\rm max} \simeq 50$ Mpc. 
Deflection by random intergalactic magnetic fields  over a distance 
$D \le D_{\rm max}$ is not capable of deflecting UHECRs in this energy 
range by more than $5-10$ degrees, given the observational limit on the 
field amplitude $B_{\rm rms}$. 

This means that the observed UHECRs in the energy range above $10^{19.5}$ eV
(some $\sim 100$ events, with about $7$ events above $10^{20}$ eV) 
must originate from multiple sources.
Solely on the basis of simple energy arguments (Rachen \& Biermann, 1993; 
Norman, Melrose  \& Achterberg, 1995), only the most luminous radio galaxies
(FRII sources) or shocks associated with ongoing large-scale ($\sim$ Mpc)
structure formation can continuously produce particles in this energy range.
Production sites in the case of radio galaxies are the so-called {\em hot spots}, 
where the impact of large-scale collimated jets on the intergalactic gas 
creates strong, large ($\sim$ kpc scale) shocks.
The typical space density of Fanaroff-Riley II radio galaxies in the local
universe is small, 
$n_{\rm s} \simeq 10^{-7} \; {\rm Mpc}^{-3}$. The corresponding 
distance scale, $D_{\rm s} = (4 \pi n_{\rm s}/3)^{-1/3} \simeq 100$ Mpc,
is large compared with the typical loss length for particles above $10^{19.5}$ eV,
$\ell(E) \simeq 10$ Mpc.
If there is still ongoing structure formation in the local universe, the
sky filling factor of the filamentary shocks associated with this process
would be small.

It is therefore not surprising that no clear candidate sources for 
UHECR events have been identified (Elbert \& Sommers, 1995).
Even if one broadens the candidate sources to include {\em all} active galaxies,
$n_{\rm s} \simeq 10^{-5} \; {\rm Mpc}^{-3}$, $D_{\rm s} \simeq 20$ Mpc,
one is hard-pressed to find a sufficient number of candidate sources within the 
local volume bounded by $D_{\rm max}$, the total number of sources scaling as
\begin{equation}
\label{contscale}
	N_{\rm s}(D \le D_{\rm max}) = 
	\left[ \frac{D_{\rm max}}{D_{\rm s}} \right]^{3} \; .
\end{equation}

A consequence of this is that the UHECR spectrum at Earth is always dominated
by a few close-by sources in the case of continuous production.
This is illustrated in figures 10 and 11.
These show two representative cases of the UHECR energy distribution, obtained
by a discrete distribution of sources within a distance of $D_{\rm in} = 250$ 
Mpc, each source distance drawn by Monte Carlo simulation 
from a distribution satisfying Eqn. (\ref{contscale}). The figure also shows 
a calculation of the background contribution due to sources at distances
from 250 Mpc to 2.5 Gpc. This background has been calculated 
using a continuous approximation for the source density.
The source density assumed equals $n_{\rm s} = 10^{-6}$ Mpc$^{-3}$, so
that $D_{\rm s} \simeq 70$ Mpc.
Particles originating at distances larger than $2.5$ Gpc will not reach 
Earth within a Hubble time due to the large accumulated delay.
These figures show that the high-energy part of the spectrum above
the GZK cut-off at $10^{19.5}$ eV is almost completely dominated by
the single closest source.

\begin{figure}%[hbtp]
\centerline{\psfig{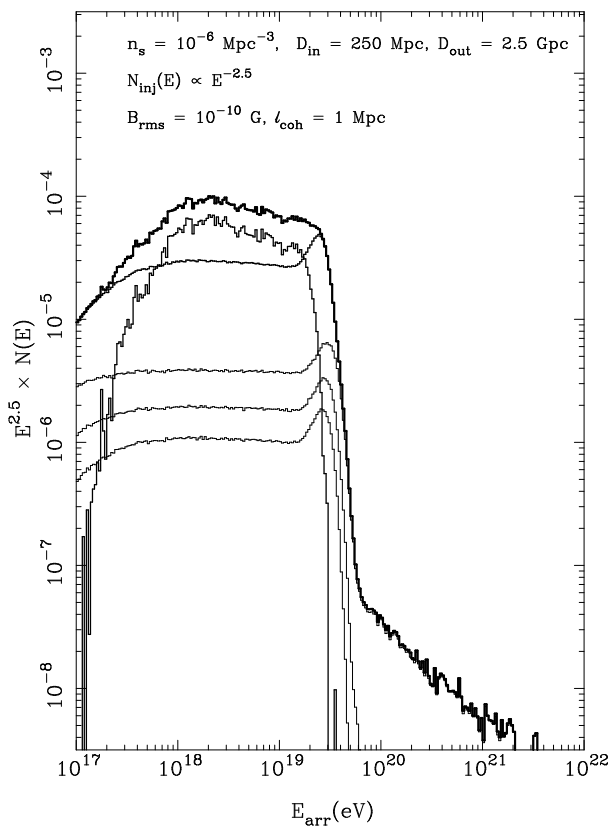}}
\caption{ The UHECR spectrum due to continuous sources with a
source density of $10^{-6}$ {\rm Mpc}$^{-3}$.
It is assumed that all sources are identical in injected flux, 
and that they are distributed  homogeneously in space so that 
$N(\le D) \propto D^{3}$.
A Monte-Carlo realization with discrete sources is used 
for $D < D_{\rm in} = 250$ {\rm Mpc}. The
background of distant sources is calculated using a continuous
distribution of source distances between 250 {\rm Mpc} and 2.5 {\rm Gpc} 
for the simulated UHECRs. 
The injection spectrum is a power-law, $N(E) = \kappa \: E^{-2.5}$.
The dominant power-law behaviour has been scaled away.
The thickest line is the total spectrum. The two thinner lines
are the contribution of the sources within 250 {\rm Mpc} and of
the background due to more distant sources, which only
contrinutes below the GZK cut-off at $10^{19.5}$ {\rm eV}.
The three thinnest curves are the spectra of the three closest sources
in this realization of the source distribution.
It is seen that the closest source completely dominates the
spectrum above the GZK cut-off. At energies below $10^{17.5} \; {\rm eV}$
the closest source again dominates the flux as particles from far-away
sources do not reach the observer within a Hubble time due to the slow
spatial diffusion of particles in this energy range.} 
\end{figure}

\begin{figure}%[hbtp]
\centerline{\psfig{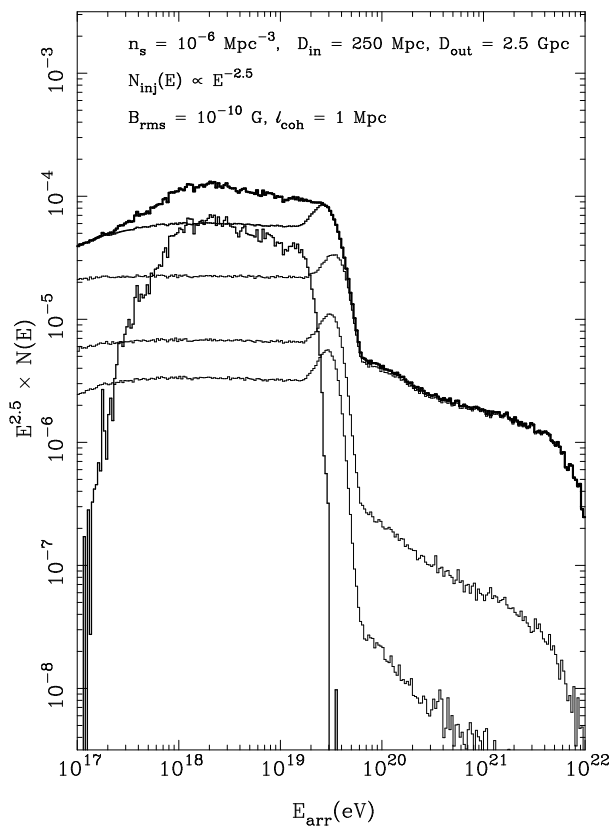}}
\caption{ As in figure 10: the UHECR spectrum due to continuous sources with the
same source density, $n_{\rm s} = 10^{-6}$ {\rm Mpc}$^{-3}$, but with a different
Monte-Carlo realization of the source distribution within a distance
$D_{\rm in} = 250$ {\rm Mpc}. In this realization 
there are a few more close sources. This distribution is compatible with
existing observations, whereas the spectrum shown in figure 10 is not.}
\end{figure} 

This simple argument lends some additional credence to the gamma-ray burster
hypothesis of  Waxman (1995a), Vietri (1995) and Milgrom \& Usov (1995),
where UHECRs are produced in bursts.

An important difference with the case where UHECRs are produced continuously is
the fact that in this case the number of sources contributing to the UHECR flux
at Earth at any given time depends critically on the time delay $t_{\rm del}(E,D)$ 
incurred during propagation: each source remains visible for a time 
$\Delta t \simeq t_{\rm del}$ as a result of the spread in arrival times.
At any one time, one sees those sources that have burst within a time interval
$\Delta t \simeq t_{\rm del}$, so that
\begin{equation}
	{\rm d} N_{\rm s} \simeq 4 \pi D^{2} \: Q_{\rm GRB} \: 
	t_{\rm del}(E,D) \: {\rm d}D
\end{equation}
sources contribute to the flux at energy $E$ from a spherical shell of 
thickness ${\rm d} D$ at distance $D$.
Here $Q_{\rm GRB}$ is the typical gamma ray burst rate per unit volume in the
cosmological scenario. The typical value of $Q_{\rm GRB}$ 
in a non-evolving scenario, where
GRBs are distributed homogeneously within a volume corresponding to a 
redshift $z \le 1$, is of order
\begin{equation}
	Q_{\rm GRB}^{\rm hom} \simeq 10^{-8} \: \left( \frac{H_{0}}
	{75 \; {\rm km} \: {\rm s}^{-1} \: {\rm Mpc}^{-1}} \right)^{-3} \;  
	{\rm Mpc}^{-3} \: {\rm yr}^{-1} \; .
\end{equation}
The total energy yield per burst is typically 
$E_{\rm GRB} \simeq 10^{51}-10^{52}$ erg, assuming a conversion efficiency
of 10 \% into gamma rays. 

However, Wijers {\em et al.} (1998) have pointed
out that most GRB formation scenarios, such as neutron star mergers
(e.g. Paczy\'nski, 1986; Narayan, Paczy\'nski \& Piran, 1992) or 
hypernovae (Paczy\'nski, 1997) involve the 
end-product of massive star evolution. One should therefore expect that the
GRB rate closely follows, on cosmological timescales, 
the star formation rate. The latter seems to peak at 
a larger redshift ($z \geq 2$). This evolving GRB scenario
implies a larger distance scale ($z \gesim 2$) and
total energy yield ($E_{\rm GRB} \sim 10^{53}$ erg) for a typical burst,
leading to a correspondingly lower burst rate per unit volume in the local 
($z \simeq 0$) universe. On the basis of the observed rate of GRBs, Wijers \etal
estimate a local burst rate equal to:
\begin{equation}
	Q_{\rm GRB}^{\rm ev} \simeq 10^{-10} \: \left( \frac{H_{0}}
	{75 \; {\rm km} \: {\rm s}^{-1} \: {\rm Mpc}^{-1}} \right)^{-3} \;  
	{\rm Mpc}^{-3} \: {\rm yr}^{-1} \; .
\end{equation}
The local burst rate is the relevant quantity for the production of UHECRs, 
which must originate within a distance $D_{\rm max}$.

The main uncertainty in these estimates is the unknown beaming factor
of the gamma radiation, which could increase the burst rate significantly
above the one derived (from observations) assuming isotropic emission.
In the hypernova scenario in particular, where the energy is generated
in a short-lived torus around a stellar core that has collapsed 
into a black hole, beaming could be significant.
Finally, given the limited number of events with X-ray, optical and/or
radio counterparts for which a distance can be derived, 
one can at this time not exclude the possibility that both the
hypernova and binary merger mechanism produce a population of GRBs.

One can define an (energy-dependent) critical distance $D_{\rm b}(E)$
which bounds the typical volume within which one expects to find 
a single source,
 $N_{\rm s}(D \le D_{\rm b}) = 1$ for given energy $E$:
\begin{equation}
	\int_{0}^{D_{\rm b}} 
	4 \pi D^{2} \: Q_{\rm GRB} \: t_{\rm del}(E,D) \: {\rm d}D = 1 \; .
\end{equation}
Using $t_{\rm del} \propto D^{2}$ it is easily checked that 
this corresponds to (Miralda-Escud\'e \& Waxman, 1996)
\begin{equation}
	\frac{4 \pi}{5} \:  Q_{\rm GRB} \: D_{\rm b}^{3} \: 
	t_{\rm del}(E,D_{\rm b}) = 1 \; .
\end{equation}
Using Eqn. (24) one finds:
\begin{equation}
	D_{\rm b}(E) \simeq 17 \: 
	\left( \frac{E_{20}}{ZB_{-9}} \right)^{2/5} 
	\: \ell_{0}^{-1/5}
	Q_{-8}^{-1/5} \; {\rm Mpc} \; ,
\end{equation}
where $Q_{-8} = Q_{\rm GRB}/(10^{-8} \; {\rm Mpc}^{-3} \: {\rm yr}^{-1})$. 
Since losses limit the source distance for $E \gesim 10^{19.5}$ eV to 
$D \leq D_{\rm max} \simeq 50$ Mpc,  one can define an energy $E_{\rm b}$
such that one expects only one source to contribute to the 
flux at that energy at any one time.
This corresponds to $D_{\rm b}(E_{\rm b}) = D_{\rm max}$ and
\begin{equation}
	E_{\rm b} \simeq 1.6 \times 10^{21} 
	ZB_{-9} Q_{-8}^{1/2} \ell_{0}^{1/2}
	\left( \frac{D_{\rm max}}{50 \; {\rm Mpc}}\right)^{5/2} 
	\; {\rm eV}
	\; .
\end{equation}
In terms of $D_{\rm b}$ and $E_{\rm b}$, the number of sources contributing 
at any time to the UHECR flux from the sphere bounded by $D_{\rm max}$ 
may be written as (Miralda-Escud\'e \& Waxman, 1996):
\begin{equation}
\label{sourcescale}
	N_{\rm s}(E,D \le D_{\rm max}) = 
	\left[ \frac{D_{\rm max}}{D_{\rm b}(E)}\right]^{5}
	 = \left[ \frac{E}{E_{\rm b}}\right]^{-2} \; .
\end{equation}
The distance $D_{\rm b}(E)$ (by definition) corresponds to the distance 
separating the local volume from which the UHECR flux at given $E$
will be intrinsically bursty and the extended
volume beyond  where the spread 
in arrival times is large and the fluxes from individual sources overlap, 
resulting in a quasi-steady UHECR flux at Earth. In a similar way, 
the energy $E_{\rm b}$ is the energy above which the UHECR flux should 
become strongly time-dependent, depending on the chance occurrence
of an UHECR production event within a distance $D_{\rm b}(E)$.

The requirement, implied by observations, 
that some $N_{\rm s} \gesim 10^{2}$ independent sources contribute to the 
UHECR flux above $E_{\rm GZK} \sim 10^{19.5}$ eV corresponds to
\begin{equation} 
	D_{\rm b}(10^{19.5} \; {\rm eV}) \lesim 
	\frac{D_{\rm max}}{100^{\frac{1}{5}}} 
	\simeq 20 \: \left( \frac{D_{\rm max}}{50 \; {\rm Mpc}} \right)
	\;  {\rm Mpc} \; .
\end{equation}
This implies a lower limit on $B_{\rm rms}$, assuming $\lc \simeq 1$ Mpc 
and protons ($Z = 1$):
\begin{equation}  
	B_{\rm rms} \gesim  2.0 \times 10^{-10} \; Q_{-8}^{-1/2} 
	\: \left( \frac{D_{\rm max}}{50 \; \rm Mpc} \right)^{-5/2}
	\; {\rm G} \; .
\end{equation}
In the non-evolving GRB scenario ($Q_{-8} \sim 1$) this is well
within the observational constraints on $B_{\rm rms}$.
These estimates correspond to a typical delay time
for particles from that volume of order
\[
	\left< t_{\rm del} \right> (10^{19.5} \; {\rm eV}, D_{\rm b}) \approx
	4.9 \times 10^{3} \: Q_{-8}^{-1} \: 
	\left( \frac{D_{\rm max}}{50 \; \rm Mpc} \right)^{-3} \; {\rm yr} 
	\; .
\]	 
If the {\em local} GRB rate per unit volume is indeed as low as proposed 
by Wijers {\em et al.} (1998) in an evolving scenario without beaming, 
$Q_{-8} \sim 10^{-2}$, the random IGM magnetic field would have to be close 
to the upper limit implied by observations: $B_{\rm rms} \simeq 10^{-9}$ G. 
In that case one has $E_{\rm b} \simeq
3 \times 10^{20}$ eV, and the most energetic cosmic ray ever observed
($E \sim 3 \times 10^{20}$ eV) is right in the energy range separating
the quasi-steady flux at lower energies and the intrinsically
bursty (intermittent) flux at higher energies.

At energies $E > E_{\rm b}$, impulsively produced particles arrive in 
bursts separated by some $2 \times 10^{2} Q_{-8}^{-1}$ yr, with the
relative burst duration scaling as $\eave{t_{\rm del}}/\tau(D_{\rm max}) =
[E/E_{\rm b}]^{-2}$. 
It is therefore unlikely that many particles
will be observed at any time at energies exceeding $\sim 10^{21}$ eV,
with $E \gg E_{\rm b}$, even if they are produced at the source
and survive in transit. This is especially true in the evolving
GRB scenario, where $E_{\rm b} \simeq 1.6 \times 10^{20} \: B_{-9} \ell_{0}$ eV.
Given that some $100$ sources contribute to the flux around $10^{19.5}$ eV,
this implies that the flux should become intrinsically bursty at an energy
$E_{\rm b} \simeq 10^{20.5}$ eV.
The observation of a single particle at $3 \times 10^{20}$ eV 
is therefore within the possibilities of this scenario.
 
Present experiments
have too small a collecting area to experimentally check the prediction 
of a quasi-steady flux below $E_{\rm b}$ and an intrinsically bursty flux 
above $E_{\rm b}$ with a `duty cycle' decaying as $E^{-2}$.
The proposed Auger observatory, which improves the available collection area
by more than an 
order of magnitude, should be able to settle this question.

The effect of the delays on the spectrum from a bursting source is shown in 
Figures 12, 13 and 14. Figure 12 shows the accumulated arrival spectrum due to a 
monochromatic  burst of particles at $E_{\rm inj} = 10^{21}$ eV for 
different integration times after the arrival of photons from the same source. 
Particles at lower energies
accumulate a longer delay, leading to a low-energy decline in the spectrum
for finite integration time.

\begin{figure}%[hbtp]
\centerline{\psfig{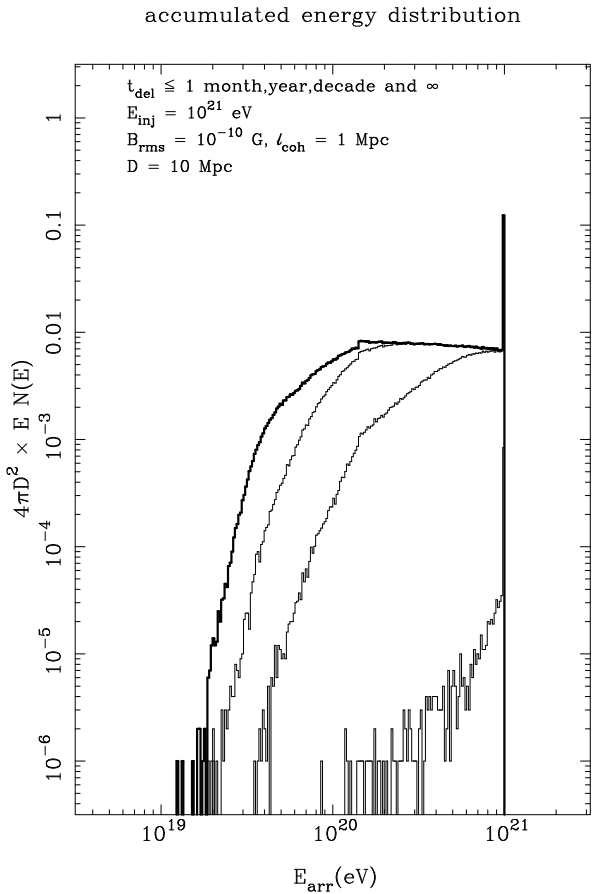}}
\caption{ Accumulated particles from a single bursting source at 10 {\rm Mpc},
as a function of the integration time after photon arrival.
The source produces particles at an energy $E_{\rm inj} = 10^{21}$ {\rm eV}.
The assumed intergalactic field has an amplitude $B_{\rm rms} = 10^{-10}$ {\rm G}. 
The time $t_{\rm del}({\rm max})$ ranges from one month to one decade. The thick line 
is the total arrival distribution $E \: N(E)$ for $t_{\rm max} \rightarrow \infty$.
For finite integration time there is a sharp decline towards low energies due 
to the strong energy dependence of the delay time, 
$\eave{t_{\rm del}} \propto E^{-2}$.
At the injection energy, the expected delay is $t_{\rm del} \simeq 0.3$ year.
The finite spread in energy of the particles is due to Poissonian statistics
of the photon encounter rate in the photo-pion production process responsible
for the energy losses from $10^{21}$ {\rm eV}.} 
\end{figure}

Figure 13 shows a similar calculation, but now for a source at a distance
of 25 Mpc injecting a
power-law spectrum $N(E) \propto E^{-2.5}$. The assumed IGM
field has an amplitude $B_{\rm rms} = 10^{-10}$ G. Shown
are the accumulated particle distributions for integration times
of 1, 10, 25 and 50 yr starting immediately
after arrival of photons from the same
source, and the total arrival spectrum for infinite integration time.

\begin{figure}%[hbtp]
\centerline{\psfig{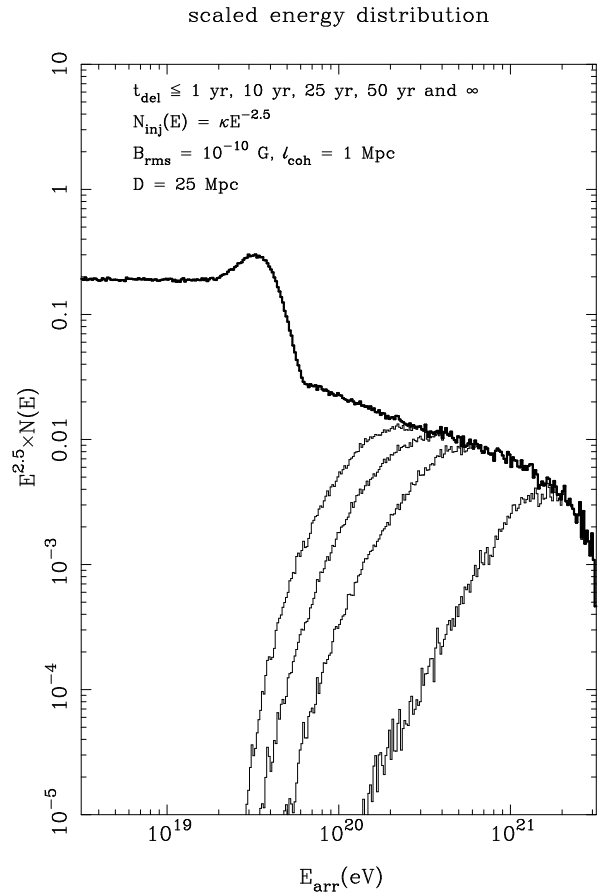}}
\caption{ The arrival energy distribution accumulated from a
source at 25 Mpc after propagation through a $10^{-10}$ {\rm G}
random field with correlation length $\lc = 1$ {\rm Mpc}.
Particles are injected with a power-law, $N(E) = \kappa\: E^{-2.5}$.
The arrival distribution is multiplied by a factor $\propto E_{\rm arr}^{2.5}$
so that an unmodified source spectrum corresponds to a horizontal line.
The integration time ranges from $1$ {\rm year} to $50$ {\rm year} after
arrival of photons from the same bursting source. 
The thick line is the total arrival
spectrum after all particles from the source have been received.}
\end{figure}

Figure 14 shows the spectrum of particles received in five
$20$-{\rm year} intervals from a source 25 Mpc distant,
with a power-law injection spectrum
$N(E) \propto E^{-2.5}$. These particles propagate through a random field
with a field strength $B_{\rm r} = 10^{-10}$ G with a coherence length
of $\lc = 1$ Mpc. Particles of lower energy are seen with 
a larger delay, with a fairly narrow spread around the mean arrival energy.
The spectrum observed at Earth consists of many such contributions from
different sources with different delays.

\begin{figure}%[hbtp]
\centerline{\psfig{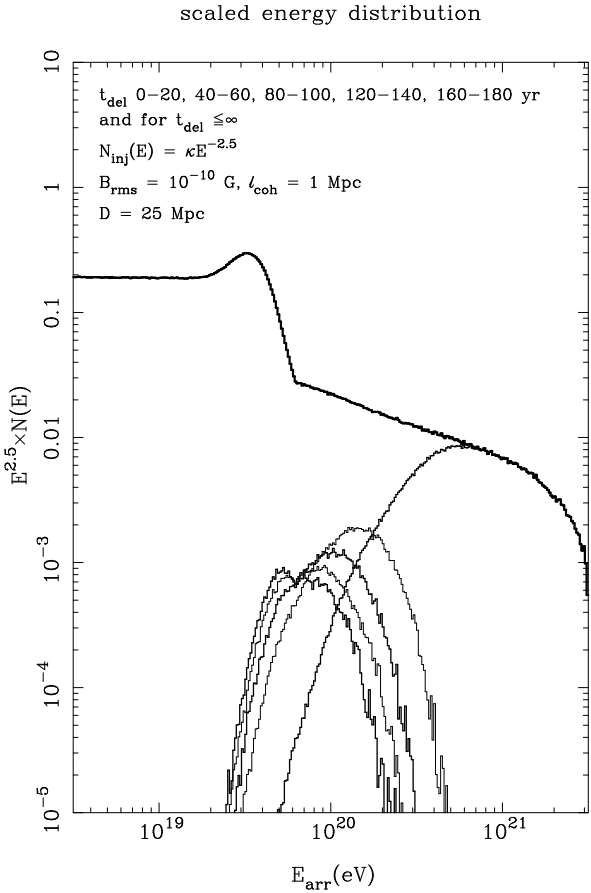}}
\caption{ Particles accumulated in five {\rm twenty-year} intervals,
equally spaced in the range $0 < \left< t_{\rm del} \right> < 180$ yr,
from a single bursting source at a distance of $25$ {\rm Mpc}, after 
propagation through a random field with field strength of $10^{-10}$ {\rm G}
and a coherence length $\lc = 1$ {\rm Mpc}.
The injection spectrum is a power-law, 
$N(E) \: {\rm d}E = \kappa \: E^{-2.5} \: {\rm d}E$.
The dominant power-law behaviour has been scaled away so that the
unmodified source spectrum is a horizontal line.
The thickest line is the total arrival spectrum which would be
accumulated over a large time interval after the arrival of
photons from the same source.}
\end{figure}

We have simulated arrival spectra that result from a distribution of
identical bursting sources which produce particles with an
energy distribution $N(E) \:{\rm d}E = \kappa E^{-s} \: {\rm d} E$.
Results are shown for $s = 2.5$. Within a distance  $D \le D_{\rm in} = 100$ Mpc,
the source distances are drawn from a distribution satisfying 
the scaling law (\ref{sourcescale}).
In addition we calculate 
the quasi-steady background from sources with a distance $D > D_{\rm in}$
up to a distance $D_{\rm max} = 2.5$ Gpc. This distance is so large that, for the
assumed field strength in the intergalactic medium, no particles reach the observer
within a Hubble time from larger distances (see Eqn. 34).
The sources at distances $D \le D_{\rm in}$ are given an `age' with respect to
the arrival of photons, distributed uniformly in the range 
$0 < t < \eave{t_{\rm del}}(E,D)$, when UHECR detection starts.
Different assumptions about source age have been tried (e.g. 
$0.5 \eave{t_{\rm del}} < t < 1.5 \eave{t_{\rm del}}$) but do not lead to
significantly different results.
The representative energies for determining source distribution and age
are chosen logarithmically equidistant,  with
$E_{\rm n} = E_{\rm b}/(\sqrt{2})^{(n-1)}$ and $n = 13$. 
Figures 15 and 16 show two representative examples of the resultant total 
spectrum for an assumed field strength $B_{\rm rms} = 10^{-10}$ G. 
They correspond to two different statistical realizations of the the source
distribution for $D \le D_{\rm in}$. We assumed a GRB rate per unit volume
of $Q_{-8} = 1$.
At energies below $E_{\ast} \sim 10^{17}$ eV, one sees the effect of 
the transition to spatial diffusion as a decline in the spectrum
towards lower energies, as discussed in Section 2.3.

\begin{figure}%[hbtp]
\centerline{\psfig{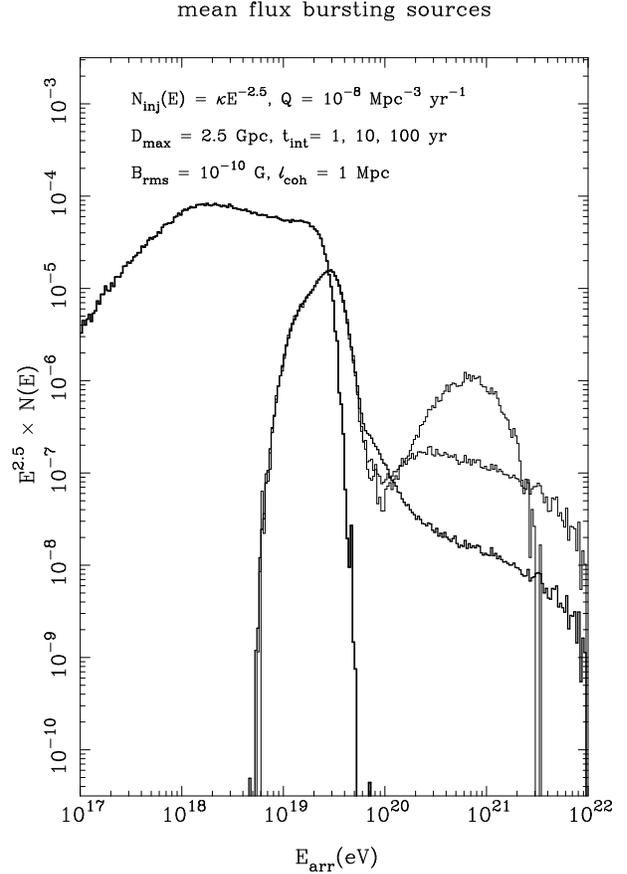}}
\caption{ The time-averaged flux$\times E^{2.5}$ due to bursting sources.
It is assumed that all sources inject an identical spectrum
$N(E) \; {\rm d}E \propto E^{-2.5} \: {\rm d}E$ for 
$10^{17} < E < 10^{22}$ {\rm eV}.
The thick line in the range $10^{17}< E_{\rm arr}<10^{19.5}$ {\rm eV} 
corresponds to the steady background due to sources more distant than 
$100$ {\rm Mpc}, the burst distance $D_{\rm b}$ at an energy
$E_{\rm b} \simeq 10^{21}$ {\rm eV} for the assumed field strength,
$B_{\rm rms} = 10^{-10}$ {\rm G}.
The decline in this contribution below $10^{18}$
{\rm eV} is due to the fact that particles undergo spatial diffusion
at these energies, and only particles from relatively nearby sources
can reach the observer within a Hubble time. The sharp cut-off at
$10^{19.5} \; {\rm eV}$ in the flux from distant sources 
is the GZK cut-off due to photo-pion production on the CMWB photons.
The high-energy part of the spectrum, which peaks around the GZK cut-off, 
is due to bursting sources within $100$ {\rm Mpc}. 
They are distributed according to the
$N(\le D) \propto D^{5}$ relation. The mean flux is shown for three
integration times: $t_{\rm int} = 1$ (thin line), $10$ (thicker line) 
and $100$ {\rm year} (thickest line). This flux varies considerably above
$10^{20}$ {\rm eV}.}
\end{figure}

\begin{figure}%[hbtp]
\centerline{\psfig{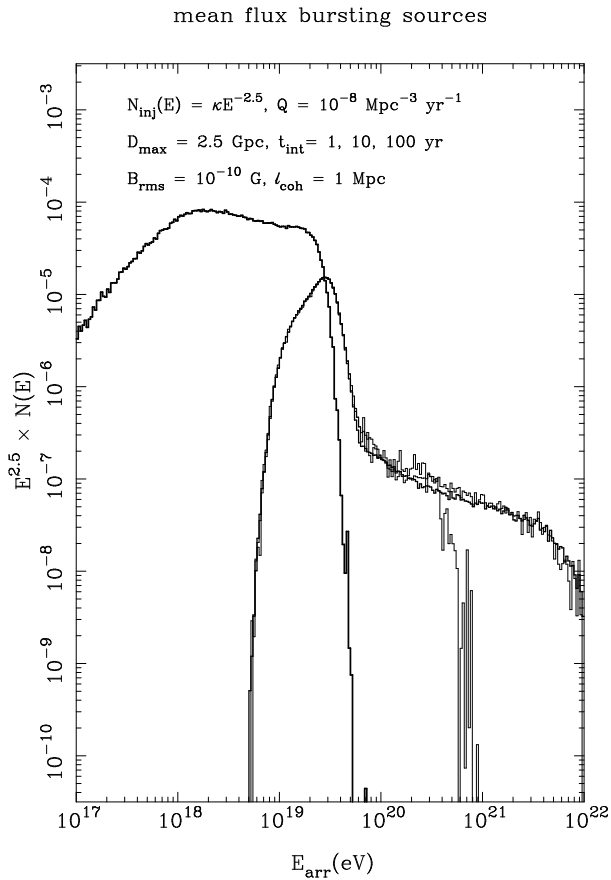}}
\caption{ As in figure 15, but with a different Monte Carlo realization of the 
source  distribution $N(\le D) \propto D^{5}$ for $D \le 100$ {\rm Mpc}.}
\end{figure}

These two examples show that there is a significant `tail' of particles
above the GZK cut-off energy. However, the flux in this tail 
depends sensitively on the observation window. Only a few sources contribute
for energies $E \le E_{\rm b}$, with the distance distribution $\propto D^{5}$
strongly favoring the largest possible distances around $D = D_{\rm max}$.
As a result, like in the case of continuous sources, the contribution
at the highest energies will be dominated by those closest few sources
which in addition have burst at the right time for the particles to reach the observer
at the present epoch.

\section{Discussion}

We have considered the propagation of ultra-high-energy cosmic rays
in a disordered but space-filling intergalactic magnetic field.
We have derived theoretical results for the accumulated deflection angle
and time delay as a function of source distance, and compared these with 
numerical simulations of the random magnetic deflection of UHECRs.
Depending on the typical magnetic field strength, the delay for particles 
originating from a distance of 50 Mpc at $10^{20}$ eV could
range from 10 years for $B = 0.01$ nG to $10^{5}$ years if $B = 1$ nG.
We found very good agreement between theory and simulations.
We have also shown that {\em spatial} diffusion due to random magnetic cells
does not occur for distances up to $100$ Mpc except
for particles with energies below $\sim 10^{19} \: B_{-9}$ eV.
Spatial diffusion does limit the background flux from distant sources
below the Greisen-Zatsepin-Kuz'min cut-off energy so that the UHECR flux
should decay below an energy of $\sim 10^{17}-10^{18}$ eV as the
sampling volume decreases with energy.

We have calculated the arrival spectra which result from the interaction of
UHECRs with the photons of the Cosmic Microwave Background Radiation,
and demonstrated that a significant flux of particles remains above
the Greisen-Zatsepin-Kuz'min cut-off energy $E_{\rm c} \sim 10^{19.5}$ eV
for source distances up to $\sim 50$ Mpc.

We discussed the implications of impulsive production of UHECRs
in gamma-ray bursts, as proposed by Waxman (1995a) and Vietri (1995). 
We have produced
composite spectra that demonstrate that, just as in the case of continuous 
production in powerful radio galaxies, it is likely that the highest energy cosmic
rays originate from only a few sources, located
at $D \lesim D_{\rm max} \simeq 50$ Mpc. These simulations imply that
the fact that the GZK cut-off around $3 \times 10^{19}$ eV is not very
pronounced in the observations is probably the result of a few UHECR
production events relatively close by. The Poisson statistics of photo-pion 
production lead to a significant high-energy tail in the particle
flux above the GZK cut-off provided enough sources are located
within 50 Mpc. The observational fact that UHECRs above the GZK cut-off energy
arrive from directions distributed over the whole sky  implies, in our view,
that the impulsive production scenario is more likely, given the low
space density of powerful radio galaxies, which seem to be the only
other likely production site of UHECRs except possibly shocks associated
with large-scale structure formation in the universe, which are equally sparse.
This conclusion of course only holds if UHECRs are hadrons and not some
exotic particle that is not subject to a significant energy loss mechanism
other than redshift as a result of the universal expansion.

In the scenario of impulsive production in gamma-ray bursts, 
the fact that $\sim 100$ events are seen above $10^{19.5}$ eV already puts a 
lower limit on the IGM magnetic field strength of $B_{\rm r} \sim 0.1-1$ nG, 
depending on the GRB rate per unit volume. For such  a field 
strength, there should not be a significant flux of cosmic rays
arriving from outside our Galaxy below $10^{17}$ eV, as slow diffusion
does not allow such particles to reach us from distant sources within
a Hubble time. Also, the typical energy scale separating the
quasi-steady flux at lower energy, due the spread in arrival times, and
the intrinsically bursty flux at higher energies is close to that
of the highest energy cosmic ray ever observed: 
$E_{\rm b} \sim 10^{20.5}-10^{21.5}$ eV.
Our simulations show that the resulting flux at Earth in this scenario 
is already strongly variable in time at an energy $E \sim 0.2 \: E_{\rm b}$. 
Therefore any interpretation of
the observed arrival spectrum in terms of a power-law fit, appropriate
for continuous sources, is in our view hazardous at high energies, as is a firm
conclusion about the presence (or absence) of a signature of the GZK 
cut-off in the observed spectrum.

We have made a significant simplification in that we assumed that the
IGM magnetic field, although randomly directed with a coherence length 
in the Mpc range, fills intergalactic space homogeneously.
A similar calculation could be done assuming the magnetic field is concentrated
mainly in clusters. In that case, one should consider the possibility
that the cluster fields are so strong ($B_{\rm cl} \sim 10^{-7}-10^{-6}$ G,
with a possible ordered component),
that true spatial diffusion of UHECRs could occur in a cluster 
for particles up to an energy of $10^{20}-10^{21}$ eV. This would lead to a
cluster-scale leaky box model, where UHECRs are confined to clusters
for a significant time (Rachen, {\em private communication}), analogous
to the situation which prevails within the Galaxy for the lower energy 
cosmic rays.  Some calculations along these lines have been
performed recently by Sigl, Lemoine and Biermann (1998) for the local 
supercluster. In that case, most of the UHECRs observed at Earth would 
be from the local group of galaxies, making the question of the production 
site even more pressing.
 
\section*{Acknowledgements}

This research was supported by the NWO-ASTRON foundation, grant nr. 781-71-050,
(AA, YG), the Research Center for Theoretical Astrophysics, University of Sydney, 
Australia (AA) and the European Union through EEC-HCM 
Contact Nr. ERBCHRX-CT94-0604 and  EU-TMR Contract Nr. ERBFMRX-CT98-0168.

\appendix
\section{Scattering on a spectrum of random fields}

We consider the general case of a broad-band 
spectrum of random turbulent magnetic fields, 
given by a Fourier expansion of the form
\begin{equation}
	\bm{B}(\bm{x}) = \int \frac{{\rm d}^{3} \bm{k}}{(2 \pi)^{3}} \:
	\tilde{\bm{B}}(\bm{k}) \: e^{\displaystyle i \bm{k} \bdot \bm{x}}
	\; ,
\end{equation}
with $\eave{\bm{B}} = 0$.
For homogeneous isotropic turbulence the Fourier components 
satisfy the relation
\begin{equation}
\label{correlat}
	\eave{ \tilde{B}_{i}(\bm{k}) \tilde{B}_{j}(\bm{k'})} = 
	\frac{{\cal B}(|\bm{k}|)}{8 \pi k^{2}} \: 
	\tilde{P}_{ij} \: 
	(2 \pi)^{6} \: \delta^{3}( \bm{k} + \bm{k'}) \; .
\end{equation}
Here we have employed the standard random-phase approximation,
and defined the projection tensor $\tilde{P}_{ij} = \delta_{ij} - \kappa_{i} \kappa_{j}$
in terms of the unit vector $\bm{\kappa} = \bm{k}/| \bm{k} |$ which
appears because the magnetic field is solenoidal: 
\begin{equation}
	\grad \bdot \bm{B} = 0 \; \; \Longleftrightarrow \; \; 
	\bm{k} \bdot \tilde{\bm{B}}(\bm{k})  = 0 \; .
\end{equation}	
The power spectrum ${\cal B}(k)$ of the magnetic field has been
normalised in such a way that
\begin{equation}
	B_{\rm rms}^{2} \equiv \eave{| \bm{B} |^{2}} = 
	\int_{0}^{\infty} {\rm d} k \: {\cal B}(k) \; ,
\end{equation} 
with $k \equiv | \bm{k} |$.

The equation of motion (\ref{eom}) for an ultra-relativistic particle
yields, by the Kubo-Taylor formula, the following expression for the 
diffusion coefficient describing the propagation direction change:
\begin{equation}
	{\cal D}_{ij} \simeq \left( \frac{Ze}{E} \right)^{2} \: 
	\int_{0}^{\infty} {\rm d}s \: \int {\rm d}k \: 
	\int \frac{{\rm d} \Omega_{k}}{8 \pi}  \: 
	{\cal B}(k) \: Q_{ij} \: 
	e^{\small{\displaystyle i \bm{k} \bdot \hatn s}} \; . 
\end{equation}
We have employed the quasi-linear approximation where the 
{\em unperturbed} motion, $\bm{x}(s) = s \: \hatn$, is used to evaluate 
the phase factor in the integral. 
To obtain this expression we have used Eqn. (\ref{correlat}), 
defined the solid angle ${\rm d} \Omega_{k}$ in Fourier space so that 
${\rm d}^{3} \bm{k} = k^{2} \: {\rm d} \Omega_{k} \: {\rm d}k$, 
and introduced the tensor
\begin{equation}
	Q_{ij} = \delta_{ij} - n_{i} n_{j} - 
	(\hatn \btimes \bm{\kappa})_{i}(\hatn \btimes \bm{\kappa})_{j}
	\; .
\end{equation}
Choosing $\hatn = \evec{z}$ and defining position angles in Fourier space by
\begin{equation}
	\bm{\kappa}  = 
	(\sin{\theta} \cos{\phi} \: , \: \sin{\theta} \sin{\phi}
	\: , \: \cos{\theta})   
\end{equation}
so that $\bm{\kappa} \bdot \hatn = \cos{\theta} \equiv \mu$, 
one can perform the integration over 
${\rm d} \Omega_{\rm k} = {\rm d} \phi \: {\rm d} \mu$ and $s$.
The integration over $\phi$ is simple, involving the integral
\begin{equation}
	\int_{0}^{2 \pi} {\rm d} \phi \: Q_{ij} =
	2 \pi \: \left( \frac{1 + \mu^{2}}{2}  \right) \: 
	\left( \delta_{ij} - n_{i} n_{j} \right) \; .
\end{equation} 
The integration over path length
$s$ only involves the complex phase factor:
\begin{equation}
	\int_{0}^{\infty} {\rm d} s \: e^{ik \mu s} = 
	\pi \: \delta(k \mu) = \frac{\pi}{k} \: \delta(\mu) \; . 
\end{equation}
The final integration over $\mu$ is now trivial, and 
the resulting expression can be written in the same form as for a 
collection of randomly oriented magnetic cells, 
\begin{equation}
	{\cal D}_{ij} = {\cal D}_{0} \: 
	\left( \delta_{ij} - n_{i} n_{j} \right) \; , 
\end{equation}
where the scalar diffusion coefficient ${\cal D}_{0}$ is given by
\begin{equation}
\label{scaldif1}
	{\cal D}_{0} = \frac{\pi}{8} \: 
	\left( \frac{Ze}{E} \right)^{2} \:
	\int_{k_{\rm min}}^{\infty} {\rm d} k \: \frac{{\cal B}(k)}{k} \; .
\end{equation}
The low wavenumber cut-off $k_{\rm min} \sim 2 \pi/\rg$ must be introduced
since cells with a size exceeding the particle gyration radius will
lead to a large-angle deflection in a single cell.
\nskip
Consider a power spectrum of the form
\begin{equation}
\label{pspec}
	{\cal B}(k) = B_{0}^{2} \lcc \: \frac{(k \lcc)^{\alpha}}
	{1 + (k \lcc)^{\alpha + \beta}} \; ,
\end{equation}
with $\alpha, \beta > 0$.
We assume that $\rg \gg \lcc$ for simplicity.
This spectrum grows as ${\cal B}(k) \propto k^{\alpha}$ for
long wavelengths ($k \lcc \ll 1$), and decays as 
${\cal B}(k) \propto k^{- \beta}$ at small scales
($k \lcc \gg 1$). Standard Kolomogorov-type MHD-turbulence, where
most energy has been fed into the turbulence at a scale $\lcc$,
predicts $\alpha \simeq 4$ and $\beta \simeq 5/3$ (e.g. Lesieur, 1987).
Substituting this in Eqn. (\ref{scaldif1}), one finds 
\begin{equation}
\label{scaldif2} 
	{\cal D}_{0} = \frac{\pi}{8} \: C_{\alpha \beta} \:
        \left( \frac{ZeB_{0}}{E} \right)^{2} \: \lcc \; ,
\end{equation}
where the constant $C_{\alpha \beta}$ is given by
\begin{equation}
	C_{\alpha \beta} = \frac{\pi}{\alpha + \beta} \: 
	{\rm cosec} \left( \frac{ \pi \alpha}{\alpha + \beta} \right) \; .
\end{equation}
This result is completely analogous with the corresponding expression
in the simple case of randomly oriented magnetic cells with identical 
field strength $B_{0}$, defining the effective coherence length as
$\lc = (3 \pi C_{\alpha \beta}/4) \lcc$. For the canonical numbers 
$\alpha = 4$ and $\beta = 5/3$ one finds $C_{\alpha \beta} \simeq 0.695$ 
and $\lc \simeq 1.64 \lcc$.

\section{Deviation angles and delays}
\setcounter{equation}{0}

The scattering of UHECRs on randomly oriented magnetic cells
can be described as a Stochastic Differential Equation (SDE)
of the It\^o form (e.g. {\O}ksendal, 1991; Gardiner, 1983).
Employing the summation convention in what follows, the stochastic equation 
of motion for the three components of the unit vector 
$\hatn = (n_{1} \: , \: n_{2} \: , \: n_{3})$ along the direction 
of flight reads:
\begin{equation}
\label{Itosde}
	{\rm d} n_{i} = - 2 {\cal D}_{0} \: n_{i} \: {\rm d}s +
	\sqrt{2 {\cal D}_{0}} \: 
	P_{ij}(\hatn) \: {\rm d} W_{j} \; .
\end{equation}
Here $P_{ij}(\hatn) \equiv \delta_{ij} - n_{i} n_{j}$ is the projection
tensor onto the plane perpendicular to $\hatn$. This tensor satisfies
$\bm{P} = \bm{P}^{\dagger}$ and $P_{ij} P_{jk} = P_{ik}$.

The quantity ${\rm d} \bm{W} \equiv ({\rm d}W_{1} \: , \: {\rm d}W_{2} \: , \: 
{\rm d}W_{3})$ is a three-component Wiener process satisfying:
\begin{equation}
\label{threewiener}
	\eave{ {\rm d}W_{i}} = 0 \; \; \; , \; \; \; 
	\eave{ {\rm d}W_{i} \: {\rm d}W_{j} } = 
	{\rm d} s \: \delta_{ij} \; . 
\end{equation}
Here the average $\eave{\cdots}$ denotes an ensemble average which
commutes with integration and differentiation with respect to the 
path length $s$. The first term in Eqn. (\ref{Itosde}) corresponds to
`dynamical friction' due to the $\hatn$-dependence of the diffusion tensor, 
and the second (stochastic) term describes true scattering.
Note that the tensor $\bm{P}$ projects ${\rm d} \bm{W}$
onto the plane perpendicular to $\hatn$, a property employed in the 
numerical realization of this process.
In the It\^o interpretation of SDEs the coefficients in a SDE are
{\em non-anticipating}, i.e. they are independent of the stochastic
increments ${\rm d} n_{i}$.

From these definitions one can immediately derive the following relation
for the stochastic integrals $n_{i}(s)$:
\begin{equation}
\label{meann}
	\frac{{\rm d} \eave{n_{i}}}{{\rm d}s} = 
	- 2 {\cal D}_{0} \: \eave{n_{i}} \; 
	\Longleftrightarrow \eave{n_{i}}(s) = \eave{n_{i}}_{0} \: 
	e^{-2 {\cal D}_{0}s} \; .
\end{equation}
We use a subscript $0$ to denote initial values at $s = 0$.
The It\^o rules for two stochastic integrals $X$ and $Y$ 
(e.g. {\O}ksendal, 1991), 
\begin{equation}
\label{Itorule}
	{\rm d}(X \cdot Y) = {\rm d}X \cdot Y + {\rm X} \cdot {\rm d}Y +
		{\rm d}X \cdot {\rm d}Y \; ,
\end{equation}
together with the relations
\begin{eqnarray}
	& & ({\rm d}s)^{2} = 0 \nonumber \\
	& & \nonumber \\
	& &  {\rm d} W_{i} \: {\rm d}s =   {\rm d} s \: 
	{\rm d}W_{i} = 0 \\
	&  & \nonumber \\
	& & {\rm d}W_{i}\: {\rm d} W_{j} = 
	{\rm d}s \: \delta_{ij} \; , \nonumber
\end{eqnarray}
lead to the following equations:
\begin{eqnarray}
\label{nsquared}
	{\rm d} \left( n_{i} \: n_{j} \right) & = &  
	2 {\cal D}_{0} \: \left( \delta_{ij} - 3\: n_{i} n_{j}
	\right) {\rm d}s +\nonumber \\
	& & \\
	& + & \sqrt{2 {\cal D}_{0}} \:
	\left\{ n_{i} P_{jk} \: {\rm d} W_{k} + 
	\left(i \rightleftharpoons j\right) 	\right\} \; .
	\nonumber
\end{eqnarray}
Taking an ensemble average one finds an equation for $\eave{n_{i} n_{j}}$:
\begin{equation}
\label{nsqave}
	\frac{{\rm d} \eave{n_{i} n_{j}}}{{\rm d}s} = 
	2 {\cal D}_{0} \: \delta_{ij} - 6 {\cal D}_{0} \: \eave{n_{i} n_{j}}
	\; .
\end{equation}
Integration yields
\begin{equation}
\label{squares}
	\eave{n_{i}n_{j}} = 
	\eave{n_{i} n_{j}}_{0} e^{-6 {\cal D}_{0}s} +
	\frac{\delta_{ij}}{3} \: \left(1 - e^{-6 {\cal D}_{0}s} \right) \; .
\end{equation}

It is easily checked that these solutions represent diffusion of the
direction of flight with diffusion coefficient
${\cal D}_{ij} = {\cal D}_{0} \: P_{ij}(\hatn)$. 
Making a first-order expansion for ${\cal D}_{0}s \ll 1$,
assuming that all members of the ensemble start with the same
(statistically sharp) initial conditions $n_{i0}$, one finds for
$\Delta n_{i} \equiv n_{i} - \eave{n_{i}}$:
\begin{equation}
	\frac{\eave{\Delta n_{i} \: \Delta n_{j}}}{2s}  
	\simeq	{\cal D}_{0} \: 
	\left( \delta_{ij} - n_{i0} n_{j0} \right)
	\equiv {\cal D}_{ij}(\hatn_{0}) \; .
\end{equation}
Next we consider the the distance integrals along the coordinate
axes,
\begin{equation}
	x_{i}(s) = \int_{0}^{s} {\rm d}s' \: n_{i}(s') \; .
\end{equation}
Taking the average, using equation (\ref{meann}), one has
\begin{equation}
\label{meanx}
	\eave{x_{i}}(s) = \frac{n_{i0}}{2 {\cal D}_{0}} \: 
	\left( 1 - e^{-2 {\cal D}_{0}s} \right) \; .
\end{equation}
From the rule (\ref{Itorule}), 
together with ${\rm d} x_{i} = n_{i} \: {\rm d}s$,
it follows that
\begin{equation}
\label{xneqn}
	{\rm d} \left( x_{i} n_{j} \right) =  
	\left( n_{i} n_{j} - 2 {\cal D}_{0} \: x_{i} n_{j} \right)  
	{\rm d} s + 
	\sqrt{2 {\cal D}_{0}} \: x_{i} P_{jk} \: {\rm d}W_{k}
	\; . 	
\end{equation}
An average yields the differential equation for $\eave{x_{i} n_{j}}$:
\begin{equation}
	\frac{{\rm d} \eave{x_{i} n_{j}}}{{\rm d}s} = 
	\eave{n_{i} n_{j}} - 2 {\cal D}_{0} \: \eave{x_{i} n_{j}} \; .
\end{equation}
Integrating this equation using Eqn. (\ref{squares}) one finds:
\begin{eqnarray}
\label{xnave}
	\eave{x_{i} n_{j}}(s) & =  &
	\frac{\displaystyle \eave{n_{i} n_{j}}_{0} - \third \: \delta_{ij}}
	{4 {\cal D}_{0}} e^{-2 {\cal D}_{0}s} \: \left(
	1 - e^{-4 {\cal D}_{0}s} \right) \nonumber \\
	& & \\
	& + & \frac{\delta_{ij}}{6 {\cal D}_{0}} \: 
	\left(1 - e^{- 2 {\cal D}_{0}s} \right) \; .
	\nonumber
\end{eqnarray}
Note that this expression is symmetric in $i$ and $j$.
In a similar fashion one can derive an equation for $\eave{x_{i} x_{j}}$:
\begin{equation}
\label{xsqeqn}
	\frac{{\rm d} \eave{x_{i} x_{j}}}{{\rm d} s} = 
	2 \: \eave{x_{i} n_{j}} \; ,	
\end{equation}
where we used the above symmetry. 
Direct integration yields:
\begin{eqnarray}
\label{xsqave}
	\eave{x_{i} x_{j}}(s) & = & 
	\frac{\displaystyle \eave{n_{i} n_{j}}_{0} - \third \: \delta_{ij}}
	{4 {\cal D}^{2}_{0}} \times \nonumber \\
	& & \nonumber \\
	& \times &
	\left[ 1 - e^{-2 {\cal D}_{0}s} - \third \: 
	\left(1 - e^{-6 {\cal D}_{0} s} \right) \right]
	\nonumber \\
	& & \\
	& + & 
	\frac{\delta_{ij}}{3 {\cal D}_{0}} \: \left[
	s - \frac{1}{2 {\cal D}_{0}} \: \left(
	1 - e^{-2 {\cal D}_{0}s} \right) \right] \; .
	\nonumber
\end{eqnarray}

These expressions can be used to calculate time delays and
deflection angles in the small-angle limit where ${\cal D}_{0} s \ll 1$.
Assuming that all particles start from the origin along the $z$-axis 
so that $\hatn(0) = (0 \: , \: 0 \: , 1)$,
one finds the following expressions to first order in ${\cal D}_{0}$:
\begin{eqnarray}
\label{smallslim}
	\eave{n_{x}} = \eave{n_{y}} = 0 & , & 
	\eave{n_{z}} = 1 - 2 {\cal D}_{0} s \; ; \nonumber \\
	& & \nonumber \\
	\eave{n_{x}^{2}} = \eave{n_{y}^{2}} = 2 {\cal D}_{0} s 
	& , & \eave{n_{z}^{2}} = 1 - 4 {\cal D}_{0} s \; ;
	\nonumber \\
	& & \nonumber \\
	\eave{x} = \eave{y} = 0 & , & 
	\eave{z} = s - {\cal D}_{0} s^{2} \; ; \nonumber \\
	& &  \\
	\eave{x^{2}} = \eave{y^{2}} = \mbox{$\frac{2}{3}$} \:
	{\cal D}_{0} s^{3}
	& , & 
	\eave{z^{2}} = s^{2} - 2 {\cal D}_{0} s^{3} \; ; \nonumber \\
	& & \nonumber \\
	\eave{xn_{x}} = \eave{yn_{y}} = {\cal D}_{0} s^{2} & , & 
	\eave{zn_{z}} = s - 3 {\cal D}_{0} s^{2} \; .
	\nonumber		
\end{eqnarray}
The distance $r$ of the particle from the origin follows, 
for $|x|, |y| \ll |z| \simeq s$,
from expanding $r = \sqrt{x^{2} + y^{2} + z^{2}}$. 
To lowest order in ${\cal D}_{0} s \ll 1$, the resulting average follows 
from the above expressions:
\begin{eqnarray}
\label{avedist}
	\eave{r}(s) & \simeq &  \eave{z} + \frac{\eave{x^{2}} + \eave{y^{2}}}{2s} 
	\nonumber \\
	& & \\
	& = & s - \mbox{$\frac{1}{3}$} \: {\cal D}_{0} \: s^{2} \; . 
	\nonumber
\end{eqnarray}	
This determines the average delay time of charged UHECRs with respect to  
ballistically propagating particles (photons) originating from the same source
at a distance $r$, to first order in ${\cal D}_{0} s$:
\begin{equation}
\label{propdelay}
	c\eave{t_{\rm del}} \simeq
	\mbox{$\frac{1}{3}$} \: {\cal D}_{0} \: r^{2} \; .
\end{equation}	
The unit vector $\bm{\hat{r}} = (x \: , \: y \: , \: z)/r$ 
along the line of sight from the source to the particle position 
$\bm{r}(s)$ at distance
$s$ makes an angle $\theta = \cos^{-1}(z/r)$ with the initial flight direction.
For symmetry reasons, assuming isotropic production of particles 
at the source, one has $\eave{\theta} = 0$. The rms value 
$\theta_{\rm rms} = \sqrt{\eave{\theta^{2}}}$ follows
from expanding $z/r \simeq 1 - (x^{2} + y^{2})/2z^{2}$ together with
$\cos{\theta} \simeq 1 - \half \theta^{2}$:
\begin{equation}
\label{thetaave}
	\eave{\theta^{2}} = \frac{\eave{x^{2}} + \eave{y^{2}}}{s^{2}} =
	\mbox{$\frac{4}{3}$} \: {\cal D}_{0} \: s \; .
\end{equation}
In a similar fashion, one can calculate the rms angle between the line 
of sight from the source to the observer and the flight direction from
\begin{equation}
	\cos{\alpha'} \equiv \hatn \bdot \bm{\hat{r}} \simeq
	\frac{n_{\rm x} x + n_{\rm y} y +  n_{\rm z} z}{s} \; .
\end{equation}
Expanding the cosine and using the above averages, one finds:
\begin{equation}
\label{barave}
	\eave{\alpha'^{2}} = 
	\mbox{$\frac{4}{3}$} \: {\cal D}_{0} \: s \; .
\end{equation}
This determines the rms angular spread in arrival directions around the
direction to the source as measured by an observer at a distance 
$r \approx s = ct$ from the source.

One can also calculate the long-time behaviour for ${\cal D}_{0} s \gg 1$.
In particular one finds that almost all memory of the initial conditions 
is lost:
\begin{eqnarray}
\label{longtime}
	\eave{n_{i}} \rightarrow 0 & , & 
	\eave{n_{i}^{2}} \rightarrow \third \; ; \nonumber \\
	& & \nonumber \\
	\eave{x} = \eave{y} = 0 & , & 
	\eave{z} \rightarrow \frac{1}{2 {\cal D}_{0}} \; ; \\
	& & \nonumber \\
	\eave{x_{i} n_{j}} \rightarrow \frac{\delta_{ij}}{6 {\cal D}_{0}} & , & 
	\eave{x_{i} x_{j}} \rightarrow \frac{s}{3 {\cal D}_{0}} \: \delta_{ij} \; .
	\nonumber
\end{eqnarray}	
The last relation corresponds to isotropic {\em spatial} diffusion, where
the distance $r$ to the source increases as	
\begin{equation}
\label{spatdiff}
	\eave{r^{2}} \simeq \frac{s}{{\cal D}_{0}} \equiv 2 \: {\cal K}t \; ,
\end{equation}
where the spatial diffusion coefficient is
\begin{equation}
\label{spatcoeff}
	{\cal K} \equiv \frac{c}{2 {\cal D}_{0}} \; .
\end{equation}
Here we use $s = ct$ for ultra-relativistic particles.	
The corresponding mean-free-path, defined by 
${\cal K} = \third \lambda_{\rm mfp}c$, is $\lambda_{\rm mfp} = 3/2{\cal D}_{0}$.

\end{document}